\documentclass[11pt]{article}

\usepackage[final]{acl}

\usepackage{times}
\usepackage{latexsym}

\usepackage[T1]{fontenc}
\usepackage[utf8]{inputenc}

\usepackage{microtype}

\usepackage{inconsolata}

\usepackage{graphicx}

\usepackage{multirow}
\usepackage{booktabs}  
\usepackage{bbding}
\usepackage{subfigure}
\usepackage{tcolorbox}
\usepackage{amsmath}
\usepackage{algorithm}
\usepackage{algorithmic}
\usepackage[table]{xcolor}
\usepackage{makecell}
\usepackage{bbold}

\title{DualGuard: Dual-stream Large Language Model Watermarking Defense against Paraphrase and Spoofing Attack}

\author{
Hao~Li$^{1,2}$, Yubing~Ren$^{1,2}$, Yanan~Cao$^{1,2\ast}$, \\
\textbf{Yingjie Li$^{1,2}$}, \textbf{Fang~Fang$^{1,2}$, Shi~Wang$^{3}$}, \textbf{Li~Guo$^{1,2}$} \\
 $^1$ Institute of Information Engineering, Chinese Academy of Sciences, Beijing, China \\
 $^2$ School of Cyber Security, University of Chinese Academy of Sciences, Beijing, China \\
 $^3$ Institute of Computing Technology, Chinese Academy of Sciences, Beijing, China \\
\texttt{lihao1998@iie.ac.cn}\\
}

\begin{document}
\maketitle

\renewcommand{\thefootnote}{\fnsymbol{footnote}}
\footnotetext[1]{Corresponding Author.}
\renewcommand{\thefootnote}{\arabic{footnote}}

\begin{abstract}
With the rapid development of cloud-based services, large language models have become increasingly accessible through various web platforms. However, this accessibility has also led to growing risks of model abuse. LLM watermarking has emerged as an effective approach to mitigate such misuse and protect intellectual property. Existing watermarking algorithms, however, primarily focus on defending against paraphrase attacks while overlooking piggyback spoofing attacks, which can inject harmful content, compromise watermark reliability, and undermine trust in attribution. To address this limitation, we propose DualGuard, the first watermarking algorithm capable of defending against both paraphrase and spoofing attacks. DualGuard employs the adaptive dual-stream watermarking mechanism, in which two complementary watermark signals are dynamically injected based on the semantic content. This design enables DualGuard not only to detect but also to trace spoofing attacks, thereby ensuring reliable and trustworthy watermark detection. Extensive experiments conducted across multiple datasets and language models demonstrate that DualGuard achieves excellent detectability, robustness, traceability, and text quality, effectively advancing the state of LLM watermarking for real-world applications.
\end{abstract}

\section{Introduction}
\label{sec introduction}
Large Language Models (LLMs) have garnered significant attention due to their capability to generate fluent, high-quality, and human-like content \cite{achiam2023gpt}. The proliferation of cloud services has further accelerated their deployment and accessibility on web platforms. Despite these advantages, the same characteristics that make LLMs attractive also exacerbate the risks of misuse, including generating malicious content \cite{gehman-etal-2020-realtoxicityprompts}, 
disseminating disinformation \cite{10.1145/3465481.3470088}, enabling impersonation \cite{ippolito-etal-2020-automatic}, and causing potential copyright infringements \cite{10.5555/3618408.3620182}. Such abuses threaten the stability and trustworthiness of the web ecosystem. To mitigate these risks, language model watermarking has recently emerged as a promising solution, aiming to embed imperceptible yet verifiable signals into LLM-generated text for reliable attribution and detection.
\begin{figure}[t]
\centering
\includegraphics[width=1.0\linewidth]{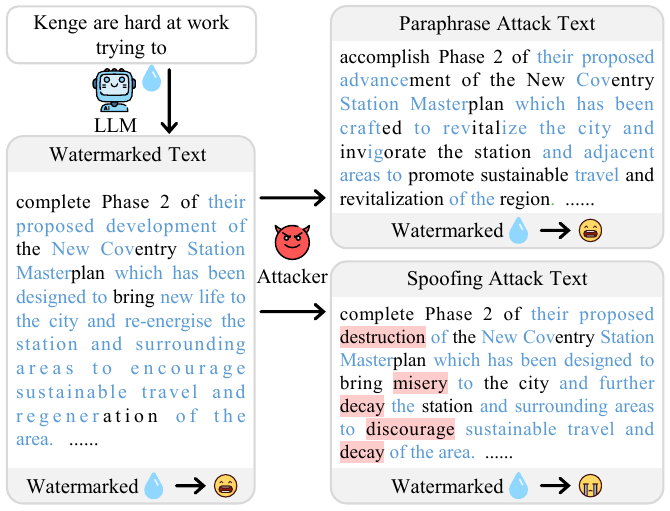}
\caption{An example is generated using the Llama-3.1-8B-Instruct model with the KGW watermarking \cite{kirchenbauer2023watermark}, where the watermark mistakenly attributes malicious content (highlighted in red) injected by the spoofing attack to the LLM.}
\label{fig: example}
\centering
\end{figure}

Existing LLM watermarking algorithms typically embed signals by modifying the logit distribution \cite{kirchenbauer2023watermark, DBLP:conf/iclr/HuCWWZH24, lu-etal-2024-entropy, lee-etal-2024-wrote, 10.5555/3692070.3694260, DBLP:conf/iclr/LiuPHM024, he-etal-2024-watermarks} or the sampling process \cite{aronsonpowerpoint, christ2024undetectable, DBLP:journals/tmlr/KuditipudiTHL24, hou-etal-2024-semstamp, hou-etal-2024-k, dathathri2024scalable}. To withstand common removal attacks, these approaches are deliberately optimized for robustness against paraphrasing attacks, ensuring that the watermark remains detectable even after semantics-preserving rewrites. However, this design choice introduces a critical vulnerability. When adversaries launch piggyback spoofing attacks \cite{DBLP:conf/nips/Pang0ZS24, an2025defending}, injecting malicious or harmful content into already watermarked text, the watermark signal often survives intact. As a result, the maliciously altered text is still flagged as “watermarked” and thus misattributed to the model. In other words, the very robustness intended to protect the model can backfire: the watermark does not safeguard the provider but instead serves as misleading evidence that falsely implicates the model in generating harmful content. This inversion of purpose highlights why defending against spoofing attacks is a fundamental requirement for trustworthy watermark deployment.

Defending against spoofing attacks presents a significant challenge, as once LLM-generated text is released, the watermark deployer has no control over or visibility into subsequent manipulations. This lack of observability leads to two critical difficulties. \textbf{First}, spoofed text often retains the original watermark signal, making it nearly indistinguishable from benign paraphrases and thereby obscuring reliable detection. \textbf{Second}, malicious content can also originate directly from the LLM itself, for example through hallucinations \cite{guerreiro-etal-2023-hallucinations} or external prompt injection \cite{gehman-etal-2020-realtoxicityprompts, 10.5555/3698900.3699003}, further blurring the line between adversarial manipulation and genuine model output. Together, these factors turn spoofing into a challenging attribution problem: defenders must not only decide whether a text carries a watermark, but also determine whether malicious content arises from the model or from external tampering. Addressing this dual challenge requires watermarking schemes that extend beyond robustness to paraphrasing and provide reliable mechanisms for detecting and tracing spoofing attacks.

To this end, we propose a novel \textbf{Dual}-stream watermarking algorithm that \textbf{Guard}s against both paraphrase and spoofing attacks by adaptively encoding two complementary watermarks (\textbf{DualGuard}), enabling accurate detection and traceability of spoofing attacks.
Our approach constructs watermark signals through the mapping model that incorporates both standard and adversarial watermark heads. The key insight is that the two watermark streams remain consistent for benign content but diverge markedly for malicious content, thereby providing a discriminative signal for the reliable detection of spoofing attacks.
During the watermark insertion stage, the algorithm selects the injected signal based on the consistency between the two watermark heads.
This adaptive mechanism allows the method to detect transitions from benign LLM-generated content to malicious spoofed content by monitoring the adversarial watermark signals. Consequently, our approach enables accurate attribution of malicious content and provides an effective means to trace spoofing attacks.
We conduct extensive experiments and in-depth analyses across multiple LLMs and datasets. The results demonstrate that our approach achieves an effective trade-off between robustness against paraphrasing and spoofing attacks, while preserving high watermark detectability.

The contributions are summarized as follows:
\begin{itemize}
    \item We explore the challenges faced by existing watermarking algorithms in defending against both paraphrase and spoofing attacks, underscoring the necessity for future research to systematically address these vulnerabilities.    
    \item We propose a novel dual-stream watermarking algorithm that is designed to defend against both paraphrase and spoofing attacks. To the best of our knowledge, this is the first watermarking scheme with the capability to reliably detect and trace spoofing attacks.
    \item We conduct extensive experiments on multiple LLMs and datasets. The results demonstrate that our method achieves an effective trade-off between robustness against paraphrasing and spoofing attacks, while preserving high watermark detectability across diverse scenarios.\footnote{Code and data are available at \url{https://github.com/hlee-top/DualGuard}.}
\end{itemize}

\begin{figure*}[t]
    \centerline{\includegraphics[width=1.0\linewidth]{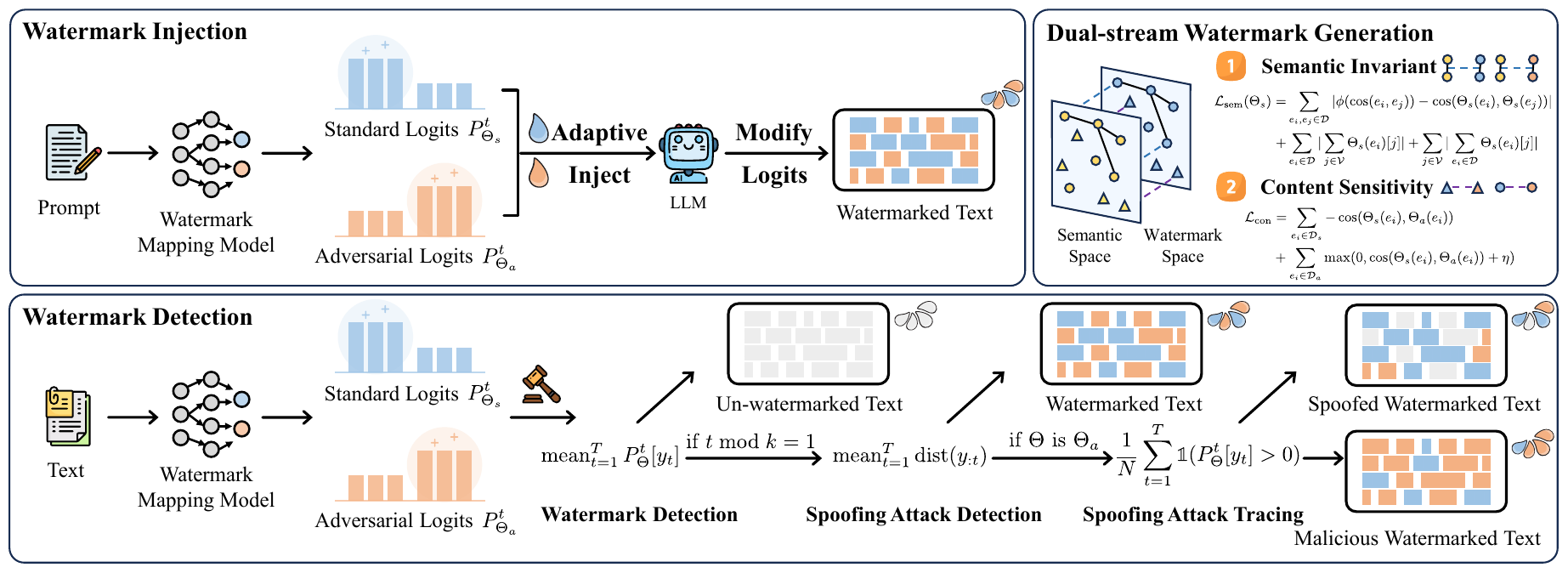}}
    \caption{Overall framework of our watermarking method DualGuard. Gray indicates un-watermarked tokens, while blue and orange denote tokens watermarked by the standard and adversarial watermark heads, respectively.}
    \label{fig: model framework}
\end{figure*}

\section{Related Works}
\subsection{Language Model Watermarking}
Language model watermarking techniques typically insert watermarks during the logits generation or token sampling process \cite{liu2024survey}. Based on the stage of insertion, these methods can be broadly categorized into two types: logits-based methods \cite{kirchenbauer2023watermark, DBLP:conf/iclr/HuCWWZH24, lu-etal-2024-entropy, lee-etal-2024-wrote, 10.5555/3692070.3694260, DBLP:conf/iclr/LiuPHM024, he-etal-2024-watermarks} and sampling-based methods \cite{aronsonpowerpoint, christ2024undetectable, DBLP:journals/tmlr/KuditipudiTHL24, hou-etal-2024-semstamp, hou-etal-2024-k, dathathri2024scalable}. Details of these methods are provided in Appendix~\ref{ap: related}.

\subsection{Piggyback Spoofing Attack}
Piggyback spoofing attacks \cite{DBLP:conf/nips/Pang0ZS24} manipulate already watermarked text by injecting malicious or harmful content, exploiting the robustness of watermarking schemes to cause the modified content to be incorrectly attributed to the original LLM \cite{gloaguen2025discovering}. Note that this attack differs from query/learning-based spoofing attacks \cite{jovanovic2024watermark}, which learn the watermarking mechanism through query or distill models to forge watermarked text (analysis in Appendix \ref{ap: security}). To defend against piggyback spoofing attacks, \citet{an2025defending} proposes a post-hoc method that trains model to remove watermark signals from text after the spoofing attack, marking the spoofed watermarked text as un-watermarked. However, this approach narrowly focuses on excluding malicious content from attribution to the watermark deployer and fails to identify malicious content, let alone trace its sources. In contrast, Dualguard represents the first watermarking algorithm capable of detecting and tracing spoofing attacks.

\section{Methodology}
In this section, we first introduce the preliminary and then present the proposed watermark mapping model for generating the dual-stream watermark signal (\S\ref{sec: model}). During text generation, these signals are iteratively injected into the LLM logits (\S\ref{sec: inject}). We then describe the watermark detection process, which encompasses both detecting the watermark signal and detecting and tracing spoofing attacks (\S\ref{sec: detect}). The overall framework is illustrated in Figure \ref{fig: model framework}, and the watermark injection and detection processes are detailed in Algorithms \ref{algo: gen} and \ref{algo: detect}.
\subsection{Preliminary}
\label{sec: preliminaries}
The generative language model $\mathcal{M}$ is defined as an autoregressive neural network with vocabulary $\mathcal{V}$. Given the current token sequence $y_{:t} = \{y_1, ..., y_{t-1}\}$,  the model $\mathcal{M}$ predicts the logit for the next token, denoted as $P_{\mathcal{M}}^t$, which are subsequently normalized via softmax function to obtain the probability distribution used for token sampling. To inject watermarks, we employ the logits-based paradigm in which additional watermark logits $P_{\Theta}^t$ are injected into the original logit: $P_{\mathcal{M}'}^t = P_\mathcal{M}^t + P_{\Theta}^t$. This adjustment biases $\mathcal{M}$ towards generating specific tokens (i.e., randomly sampled green list tokens in KGW \cite{kirchenbauer2023watermark}). 
During detection, the frequency of these tokens in the generated sequence is evaluated, and the statistical test is applied to determine the presence of the watermark.

\subsection{Dual-stream Watermark Generation}
\label{sec: model}
Watermark generation is a critical component of watermarking algorithms. Existing approaches employ techniques such as entropy \cite{lu-etal-2024-entropy, lee-etal-2024-wrote} and semantic invariant watermarks \cite{DBLP:conf/iclr/LiuPHM024, he-etal-2024-watermarks, 10.5555/3692070.3693308} to improve robustness against paraphrase attacks and ensure watermark detectability even after textual alterations. However, these designs introduce a critical vulnerability. When adversaries launch piggyback spoofing attacks by injecting malicious or harmful content into already watermarked text, the watermark signal often remains intact. Consequently, the maliciously altered text is still labeled as watermarked and is misattributed to the model. Such misattribution severely undermines both the reliability of the watermark scheme and the reputation of the watermark deployer.

To address this challenge, we propose an adaptive dual-stream watermarking algorithm, called DualGuard. The core idea is to adaptively inject standard and adversarial watermarks according to the semantics of the content generated by the LLM. These two watermark signals remain consistent for benign text but diverge for malicious text, a property that enables reliable detection of spoofing attacks. Furthermore, by iteratively applying the standard watermark to benign text and the adversarial watermark to malicious text, DualGuard effectively captures the transition from benign to malicious content under spoofing attacks (i.e., when the standard watermark head is converted by the adversarial watermark head), thereby enabling accurate tracing of spoofing attacks.

Specifically, we train the watermark mapping model $\mathcal{G}$ to generate the dual-stream watermark signal. The model consists of the shared multi-layer feed-forward neural network with residual connections, along with the standard watermark head $\Theta_s$ and the adversarial watermark head $\Theta_a$:
\begin{align*}
\Theta_s(e_t), \Theta_a(e_t) = \mathcal{G}(e_t),
\end{align*}
where $\Theta_s(e_t)$ and $\Theta_a(e_t)$ denote the outputs of the standard and adversarial watermark heads at time step $t$, respectively. 
$e_t = \mathcal{E}(y_{t-\rho:t})$ denotes the embedding of the current token subsequence $y_{t-\rho:t}$ obtained through the encoding model $\mathcal{E}$, where $\rho$ represents the watermark prefix length. The watermark mapping model $\mathcal{G}$ is designed to ensure that the resulting dual-stream watermark signal satisfies the following properties:

\paragraph{Semantic Invariant} Semantic invariant watermarks possess three essential characteristics \cite{DBLP:conf/iclr/LiuPHM024, he-etal-2024-watermarks, 10.5555/3692070.3693308}. First, similar texts should produce similar watermark signals, thereby ensuring robustness against minor modifications such as paraphrasing. Second, the watermark signal must perturb the vocabulary in a balanced manner, such that the number of tokens with increased probabilities equals the number with decreased probabilities, i.e., the entries in watermark logits contain an equal number of positive and negative values. Finally, the watermark should remain unbiased with respect to the vocabulary, introducing no statistical preference for specific tokens and thereby preserving the generative distribution of the model. To this end, we formulate and minimize the semantic loss $\mathcal{L}_{\text{sem}} = \mathcal{L}_{\text{sem}}(\Theta_s) + \mathcal{L}_{\text{sem}}(\Theta_a)$, where $\mathcal{L}_{\text{sem}}(\Theta_s) = $
\begin{align*}
&\sum_{{e_i,e_j} \in \mathcal{D}}|\phi(\cos(e_i, e_j))  -  \cos(\Theta_s(e_i), \Theta_s(e_j))|  \\
&+ \sum_{e_i \in \mathcal{D}}|\sum_{j \in \mathcal{V}}\Theta_s(e_i)[j]| + \sum_{j \in \mathcal{V}}|\sum_{e_i \in \mathcal{D}}\Theta_s(e_i)[j]|,
\end{align*}
where $\mathcal{L}_{\text{sem}}(\Theta_s)$ and $\mathcal{L}_{\text{sem}}(\Theta_a)$ represent the semantic losses of the dual-stream watermark heads, $\mathcal{D}$ means the watermark mapping model training dataset, $\cos$ represents cosine similarity, and $\phi(x) = \tanh(\tau(x - \bar{x}))$ is the scaling function based on the mean cosine similarity of the original data, which makes similar embeddings generate
more relevant watermark signals, and vice versa.

\paragraph{Content Sensitivity} The dual-stream watermark signal is inherently content-sensitive, remaining consistent for benign content while exhibiting significant divergence for malicious content. This property enables our method to detect potential spoofing attacks in the streaming manner and to adaptively inject corresponding watermark signals during text generation, thereby providing an effective defense against such attacks. To this end, we formulate and minimize the contrastive loss:
\begin{align*}
& \mathcal{L}_{\text{con}} = \sum_{e_i \in \mathcal{D}_s} -\cos(\Theta_s(e_i), \Theta_a(e_i)) \nonumber \\
& + \sum_{e_i \in \mathcal{D}_a}\max(0, \cos(\Theta_s(e_i), \Theta_a(e_i))+\eta),
\end{align*}
where $\mathcal{D}_s$ and $\mathcal{D}_a$ represent the benign and malicious text subsets in the training set $\mathcal{D} = \mathcal{D}_s  \cup \mathcal{D}_a$, $\eta$ means the separation margin hyperparameter.

Considering all the properties, the loss function of the watermark mapping model is:
\begin{align*}
    \mathcal{L} = \mathcal{L}_{\text{sem}} + \lambda\mathcal{L}_{\text{con}}.
\end{align*}

\subsection{Watermark Injection}
\label{sec: inject}
We leverage the content sensitivity of the watermark mapping model to adaptively inject dual-stream watermarks. Standard watermark head is applied to benign content, whereas adversarial watermark head is applied to malicious content. Under spoofing attacks, the LLM-generated text is maliciously altered. Since benign and malicious content employ different watermark heads, this dual-stream design effectively captures such transformations, thereby enabling reliable tracing of spoofing attacks. The overall process of watermark text generation is summarized in Algorithm \ref{algo: gen}.

Specifically, the generated token sequence $y_t$ is divided into fixed-length windows. By default, the standard watermark head $\Theta_s$ is applied. When the current token $y_t$ corresponds to the beginning of a new window ($t\bmod k = 1$), the watermark head is selected based on the current generated content: 
\begin{align*}
& {\Theta}  = 
    \begin{cases}
        {\Theta}_s,  \operatorname{dist}(y_{:t}) < \alpha \\
        {\Theta}_a,  \operatorname{dist}(y_{:t}) \geq \alpha 
    \end{cases}, \text{if } t \bmod k=1,  \\
& \operatorname{dist}(y_{:t}) = 1- \cos({\Theta}_s(\mathcal{E}(y_{:t})), {\Theta}_a(\mathcal{E}(y_{:t}))),
\end{align*}
where $\Theta$ denotes the watermark head selected for token sequence $y_{:t}$, $\alpha$ is the selection threshold, $k$ represents the window length, and $\operatorname{dist}(y_{:t})$ denotes the cosine distance between the outputs of the standard watermark head $\Theta_s$ and the adversarial watermark head $\Theta_a$ on the token sequence $y_{:t}$.

Then, the output of the selected watermark head is scaled and mapped to the dimensionality of the LLM vocabulary $\mathcal{V}$ to obtain final watermarked logits \cite{DBLP:conf/iclr/LiuPHM024}. This design ensures that our method can be seamlessly applied to LLMs with different vocabularies. Formally, the watermarked logits at time step $t$ are computed as:
\begin{align*}
P_\Theta^t = \operatorname{F}(\tanh(\gamma {\Theta}(e_{t}))),
\end{align*}
where $\gamma$ means the scaling factor, the watermark logits are scaled to be close to 1 or -1 after the tanh function. $\operatorname{F}(\cdot)$ denotes the mapping function that randomly projects the output dimension of the watermark head $\Theta$ onto the LLM vocabulary, i.e., the watermark head output is repeatedly mapped to the higher-dimensional vocabulary. Finally, the watermark logits are injected into original LLM logits via scaling parameter $\delta$, which proportionally amplifies or suppresses original logits and thereby minimizes the impact of perturbations on original probability distribution \cite{10.5555/3692070.3693308}:
\begin{align*}
    P_{\mathcal{M}'}^t = P_{\mathcal{M}}^t + \delta \cdot P_{\mathcal{M}}^t P_{\Theta}^t.
\end{align*}

\subsection{Watermark Detection}
\label{sec: detect}
With the dual-stream watermarking mechanism, DualGuard generates stable and discriminative watermark signals for watermark detection, spoofing attack detection, and spoofing attack tracing. The complete process is shown in Algorithm \ref{algo: detect}.

\paragraph{Watermark Detection} 
We formulate the null hypothesis: the candidate text is not watermarked. If the average watermark logit value across all tokens exceeds zero, the null hypothesis is rejected, indicating that the candidate text is watermarked. During the injection stage, both the standard and adversarial watermark heads are adaptively applied. Consequently, the watermark head $\Theta$ is first determined using the window $k$, after which the watermark logit values of all tokens are computed:
\begin{align*}
    \text{Score}_{\text{wd}} = \operatorname{mean}_{t=1}^{T} P_{\Theta}^t[y_t].
\end{align*}
\paragraph{Spoofing Attack Detection} 
When subjected to spoofing attacks, the text still retains a strong watermark signal, leading existing watermarking algorithms to mistakenly attribute the injected malicious content to the LLM itself and thereby compromising reliable detection.
Benefiting from the content sensitivity property of watermark signal, we detect spoofing attacks based on this feature:
\begin{align*}
\text{Score}_{\text{sd}} = \operatorname{mean}_{t=1}^{T}\operatorname{dist}(y_{:t}), \text{if } t \bmod k=1, 
\end{align*}
where output $y_{:t-1}$ that are potentially affected by spoofing attacks receive higher scores, since the dual-stream watermark heads produce significantly different logits for maliciously modified text compared with benign content.

\paragraph{Spoofing Attack Tracing} Current LLMs employ various safeguards to mitigate malicious content generation. However, these defenses are frequently bypassed in real-world scenarios, leading to harmful outputs such as hallucinations \cite{guerreiro-etal-2023-hallucinations} and prompt injection attacks \cite{gehman-etal-2020-realtoxicityprompts, 10.5555/3698900.3699003}. This complicates defenses against spoofing attacks, as both malicious content directly generated by LLMs and spoofed texts produced by attackers may carry valid watermark signals alongside malicious or harmful content.

\begin{table*}[ht]
\centering
\resizebox{\linewidth}{!}{
    \begin{tabular}{lcccccccccccccc}
    \toprule
    \multirow[c]{3}{*}{\textbf{Method}} & \multicolumn{7}{c}{\textbf{RealNewsLike}} & \multicolumn{7}{c}{\textbf{BookSum}} \\
    \cmidrule(lr){2-8}  \cmidrule(lr){9-15}
    & \multicolumn{3}{c}{$\textbf{Robustness}_{\textbf{para}}$} &  \multicolumn{3}{c}{$\textbf{Robustness}_{\textbf{spoof}}$} & \multirow{2}{*}{\makecell{\textbf{Overall}\\ \textbf{AUC}}}
    & \multicolumn{3}{c}{$\textbf{Robustness}_{\textbf{para}}$} & \multicolumn{3}{c}{$\textbf{Robustness}_{\textbf{spoof}}$} 
    & \multirow{2}{*}{\makecell{\textbf{Overall}\\ \textbf{AUC}}} \\
    \cmidrule(lr){2-4}  \cmidrule(lr){5-7}     \cmidrule(lr){9-11}  \cmidrule(lr){12-14}

    & \textbf{AUC} & \textbf{TP@5\%} & \textbf{TP@10\%} & \textbf{AUC} & \textbf{TP@5\%} & \textbf{TP@10\%} & & \textbf{AUC} & \textbf{TP@5\%} & \textbf{TP@10\%} & \textbf{AUC} & \textbf{TP@5\%} & \textbf{TP@10\%} & \\
    \midrule
    \rowcolor[gray]{0.9} \multicolumn{15}{c}{OPT-1.3B} \\

    \textbf{KGW} & 0.9871 & 0.9150 & 0.9600 & 0.5141 & 0.0667 & 0.0923 & 0.7506 & 0.9777 & 0.9000  & 0.9300 & 0.4613 & 0.0611 & 0.0833 & 0.7195 \\
    \textbf{Unbiased} & 0.5011 & 0.0496 & 0.0993 & 0.4956 & 0.0492 & 0.0985 & 0.4983 & 0.5162 & 0.0530 & 0.1060 & 0.4729 & 0.0455 & 0.0911 & 0.4945 \\
    \textbf{AAR} & 0.7513 & 0.0700 & 0.1850 & 0.3785 & 0.0160 & 0.0588 & 0.5649 & 0.7834 & 0.1000 & 0.2800 & 0.5308 & 0.0977 & 0.1782 & 0.6571 \\
    \textbf{SynthID} & 0.7108 & 0.2050 & 0.2900 & 0.5786 & 0.0622 & 0.2021 & 0.6447 & 0.7559 & 0.1950 & 0.3250 & 0.4500 & 0.0543 & 0.1304 & 0.6029 \\
    \textbf{EWD} & 0.9759 & 0.8950 & 0.9350 & 0.5780 & 0.0990 & 0.1927 & 0.7769 & 0.9821 & 0.9400 & 0.9700 & 0.4733 & 0.0278 & 0.0667 & 0.7277 \\
    \textbf{SWEET} & 0.9731 & 0.8550 & 0.9350 & 0.5730 & 0.1031 & 0.2113 & 0.7730 & 0.9849 & 0.9250 & 0.9700 & 0.5136 & 0.0798 & 0.1117 & 0.7492 \\
    \textbf{DIPmark} & 0.5161 & 0.1150 & 0.1650 & 0.5569 & 0.0729 & 0.1354 & 0.5365 & 0.5351 & 0.0550 & 0.1750 & 0.5375 & 0.0973 &  0.1459 & 0.5363 \\
    \textbf{SIR} & 0.9235 & 0.6050 & 0.8100 & 0.4190 & 0.0308 & 0.0667 & 0.6713 & 0.9306 & 0.7050 & 0.7950 & 0.4190 & 0.0330 & 0.0659  & 0.6748 \\
    \textbf{XSIR} & 0.9224 & 0.6250 & 0.7400 & 0.4300 & 0.0306 & 0.0561 & 0.6762 & 0.9601 & 0.7900 & 0.8900 & 0.3882 & 0.0108 & 0.0649 & 0.6741 \\
    \textbf{DualGuard} & 0.9680 & 0.8600 & 0.9250 & 0.9284 & 0.3505 & 0.8247 & \textbf{0.9482} & 0.9760 & 0.9200 & 0.9550 & 0.9552 & 0.7784 & 0.8693 & \textbf{0.9656} \\
    \midrule
    \rowcolor[gray]{0.9} \multicolumn{15}{c}{Llama3.1-8B-Instruct} \\
    \textbf{KGW} & 0.8734 & 0.5050 & 0.6600 & 0.5573 & 0.1451 & 0.2124 & 0.7153 & 0.8999 & 0.6850 & 0.7350 & 0.4842 & 0.0688 & 0.0813 & 0.6920 \\
    \textbf{Unbiased} & 0.5003 & 0.0516 & 0.1033 & 0.5021 & 0.0500 & 0.1000 & 0.5012 & 0.5107 & 0.0517 & 0.1033 & 0.4914 & 0.0484 & 0.0968 & 0.5010 \\
    \textbf{AAR} & 0.7117 & 0.1400 & 0.2450 & 0.4332 & 0.0410 & 0.0769 & 0.5724 & 0.7329 & 0.2050 & 0.3350 & 0.5073 & 0.0403 & 0.1074 & 0.6201 \\
    \textbf{SynthID} & 0.6139  & 0.0500 & 0.1100  & 0.5686  & 0.0838  & 0.1571  & 0.5912 & 0.6686 & 0.0900  & 0.2000  & 0.4941 & 0.0390 & 0.0649 & 0.5813 \\
    \textbf{EWD} & 0.9410 & 0.7800 & 0.8600 & 0.5031 & 0.1077 & 0.1590 & 0.7220 & 0.9270  & 0.7500  & 0.8050 & 0.4354 & 0.0496 & 0.0780 & 0.6812 \\
    \textbf{SWEET} & 0.9185 & 0.6250 & 0.7450 & 0.5564 & 0.0208 & 0.1354 & 0.7374 & 0.9040 & 0.6500 & 0.7650 & 0.4269 & 0.0200 & 0.0667 & 0.6654 \\
    \textbf{DIPmark} & 0.5337 & 0.1250 & 0.1900 & 0.5123 & 0.0263 & 0.0632 & 0.5230 & 0.5512 & 0.0900 & 0.1700 & 0.5006 & 0.0845 & 0.1197 & 0.5259 \\
    \textbf{SIR} & 0.9274 & 0.6900 & 0.7850 & 0.4466 & 0.0052 & 0.0309 & 0.6870 & 0.8026 & 0.3050 & 0.4400 & 0.3378 & 0.0312 & 0.0563 & 0.5702 \\
    \textbf{XSIR} & 0.7968 & 0.4500 & 0.5250 & 0.5069 & 0.0695 & 0.0909 & 0.6518 & 0.8709 & 0.5300 & 0.6250 & 0.4921 & 0.0645 & 0.0903 & 0.6815 \\
    \textbf{DualGuard} & 0.9244 & 0.6200 & 0.7600 & 0.9159 & 0.2552 & 0.6562 & \textbf{0.9201} & 0.9253 & 0.6450 & 0.8050 & 0.9354 & 0.5655 & 0.7448 & \textbf{0.9303} \\
    \bottomrule
    \end{tabular}
    }
\caption{Experimental results of paraphrase attack robustness ($\text{Robustness}_{\text{para}}$) and spoofing attack robustness ($\text{Robustness}_{\text{spoof}}$) on the RealNewsLike and BookSum datasets, with watermark detectability presented in Figure \ref{fig: watermark detect}.}
\label{tab: main result}
\end{table*}

To address this challenge, DualGuard assigns distinct watermark heads to benign and malicious content. For malicious content generated by LLM, the adversarial watermark head is applied during both injection and detection, leading to a higher proportion of watermark tokens generated when the adversarial watermark head is selected. In contrast, for text subjected to spoofing attacks, the original benign text is watermarked with the standard watermark head. When such text is later modified into malicious content through spoofing, DualGuard adaptively selects the adversarial watermark head during detection, resulting in a lower proportion of watermark tokens produced under the adversarial head. By exploiting this asymmetry, DualGuard exhibits differentiated behaviors across the two types of malicious content, thereby enabling the reliable detection of benign-to-malicious transformations induced by spoofing attacks and facilitating precise source tracing of malicious content through the adversarial watermark head:
\begin{align*}
\text{Score}_{\text{st}} =  \frac{1}{N}\sum_{t=1}^{T} \mathbb{1}(P_{\Theta}^t[y_t] > 0), \text{if } \Theta \text{ is } {\Theta}_a,
\end{align*}
where $\mathbb{1}$ denotes the indicator function, and $N$ is the total number of tokens generated under the adversarial watermark head. $\text{Score}_{\text{st}}$ measures the proportion of adversarial watermark tokens produced under the adversarial watermark head ${\Theta}_a$.

\section{Experiments}
\subsection{Experimental Settings}
\label{sec exp}
\paragraph{Datasets}
We conduct experiments on the RealNewsLike subset of C4 \cite{10.5555/3455716.3455856}, BookSum \cite{kryscinski-etal-2022-booksum}, RealToxicityPrompts \cite{gehman-etal-2020-realtoxicityprompts}, and RTP-LX \cite{de2025rtp} datasets. Detailed data analysis and processing are listed in Appendix \ref{ap: dataset}.

\paragraph{Paraphrase and Spoofing Attack details} 
Paraphrase attacks aim to rephrase the original text while preserving its semantics, whereas spoofing attacks maliciously alter the text by injecting malicious or harmful content. The prompts used in our experiments are shown in Figures \ref{fig: paraphrase prompt} and \ref{fig: spoof prompt}, with GPT-4.1 serving as the underlying LLM. To ensure the effectiveness of spoofing attacks, we employ a binary classification detector to filter spoofed text as malicious, with the detection threshold set to 0.5. Additional details of the attack setup and analysis are provided in Appendix \ref{ap: prompt} and \ref{ap: spoof analysis}. 
\paragraph{Evaluation Metrics}
We employ the following metrics to evaluate watermarking algorithms:
Watermark Detectability, Paraphrase Attack Robustness, Spoofing Attack Robustness, and Traceability. Details of these metrics are listed in Appendix~\ref{ap: metrics}.

\begin{table*}[ht]
\centering
\resizebox{\linewidth}{!}{
    \begin{tabular}{lcccccccccccccc}
    \toprule
    \multirow{3}{*}{\textbf{Method}} & \multicolumn{7}{c}{\textbf{RealToxicityPrompts}} & \multicolumn{7}{c}{\textbf{RTP-LX}} \\
    \cmidrule(lr){2-8}  \cmidrule(lr){9-15}
    &  \multicolumn{3}{c}{\textbf{OPT-1.3B}} & \multicolumn{3}{c}{\textbf{Llama3.1-8B-Instruct}} 
    & \multirow{2}{*}{\makecell{\textbf{Overall}\\ \textbf{AUC}}}
    &  \multicolumn{3}{c}{\textbf{OPT-1.3B}} & \multicolumn{3}{c}{\textbf{Llama3.1-8B-Instruct}} 
    & \multirow{2}{*}{\makecell{\textbf{Overall}\\ \textbf{AUC}}} \\
    \cmidrule(lr){2-4}   \cmidrule(lr){5-7} \cmidrule(lr){9-11} \cmidrule(lr){12-14}
    & \textbf{AUC} & \textbf{TP@5\%} & \textbf{TP@10\%}  & \textbf{AUC} & \textbf{TP@5\%} & \textbf{TP@10\%} & & \textbf{AUC} & \textbf{TP@5\%} & \textbf{TP@10\%} & \textbf{AUC} & \textbf{TP@5\%} & \textbf{TP@10\%} & \\
    \midrule
    \textbf{KGW} & 0.4454 & 0.0350 & 0.0500 & 0.5541 & 0.0500 & 0.1550 & 0.4997 & 0.5057 & 0.0200 & 0.0450 & 0.5848 & 0.0600 & 0.1500 & 0.5452 \\
    \textbf{Unbiased} & 0.5462 &  0.0588 & 0.1175 & 0.5379 & 0.0617 & 0.1233 & 0.5421 & 0.4997 & 0.0512 & 0.1024 & 0.4817 & 0.0509 & 0.1018 & 0.4907 \\
    \textbf{AAR} & 0.5918 & 0.1050 & 0.1750 & 0.4940 & 0.0500 &  0.0900 & 0.5429 & 0.5333 & 0.0800 & 0.1150 & 0.4265 & 0.0750 & 0.1050 & 0.4799 \\
    \textbf{SynthID} & 0.4166 & 0.0150 & 0.0900 & 0.5465 & 0.0850 & 0.1800 & 0.4815 & 0.4701 & 0.0450 & 0.0550 & 0.5641 & 0.0800 & 0.2100 & 0.5171 \\
    \textbf{EWD} & 0.4190 & 0.0250 & 0.0450 & 0.5961 & 0.0650 & 0.1550 & 0.5075 & 0.5289 & 0.0300 & 0.0800 & 0.6080 & 0.0650 & 0.1350 & 0.5684 \\
    \textbf{SWEET} & 0.4188 & 0.0150 & 0.0450 & 0.5451 & 0.0550 & 0.1150 & 0.4819 & 0.4965 & 0.0400 & 0.0700 & 0.5874 & 0.1000 & 0.1600 & 0.5419 \\
    \textbf{DIPmark} & 0.4948 & 0.0500 & 0.0850 & 0.4539 & 0.0101 & 0.0354 & 0.4743 & 0.5185 & 0.0650 & 0.0950 & 0.4775 & 0.0415 & 0.0622 & 0.4980 \\
    \textbf{SIR} & 0.5336 & 0.0950 & 0.2000 & 0.6445 & 0.2250 & 0.3050 & 0.5890 & 0.6327 & 0.1050 & 0.1800 & 0.7225 & 0.1313 & 0.3030 & 0.6776 \\
    \textbf{XSIR} & 0.5114 & 0.0450 & 0.1050 & 0.5441 & 0.0950 & 0.1450 & 0.5277 & 0.5116 & 0.0750 & 0.1450 & 0.5820 & 0.0833 & 0.1667 & 0.5468 \\
    \textbf{DualGuard} & 0.9011 & 0.5200 & 0.6600 & 0.8513 & 0.3750 & 0.5700 & \textbf{0.8762} & 0.8704 & 0.5100 & 0.6500 & 0.8497 & 0.3800 & 0.5550 & \textbf{0.8600} \\
    \bottomrule
    \end{tabular}
}
\caption{Experimental results of spoofing attack traceability on RealToxicityPrompts and RTP-LX dataset.}
\label{tab: traceability}
\end{table*}

\begin{figure*}[h]
\centering
\subfigure[Gemini-2.5]{
\begin{minipage}[t]{0.5\linewidth}
\centering
\includegraphics[width=\linewidth]{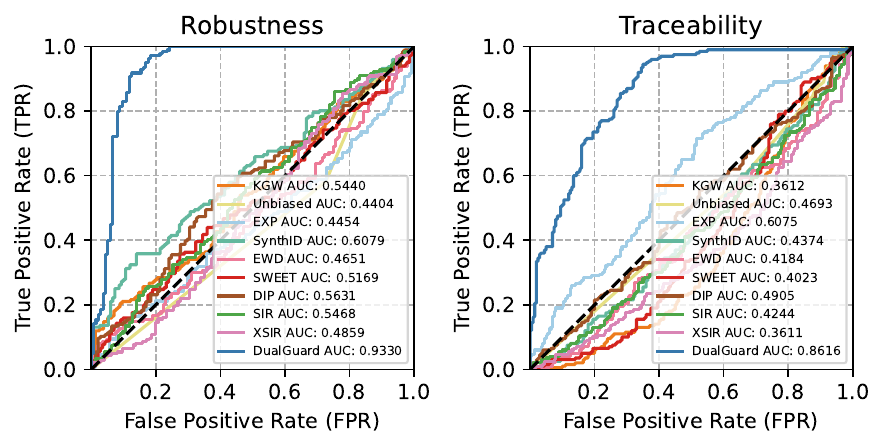}
\end{minipage}%
}%
\subfigure[Qwen3]{
\begin{minipage}[t]{0.5\linewidth}
\centering
\includegraphics[width=\linewidth]{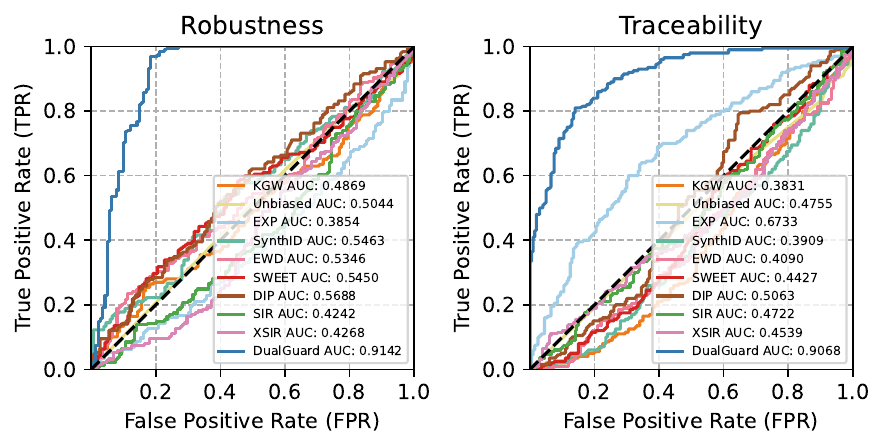}
\end{minipage}%
}%
\caption{Experimental results of different attack models on RealNewsLike and RealToxicityPrompts dataset.}
\label{fig: other attack}
\end{figure*}

\paragraph{Baselines} We evaluate our method against the following representative watermarking algorithms, including logits-based methods  KGW \cite{kirchenbauer2023watermark}, Unbiased \cite{DBLP:conf/iclr/HuCWWZH24}, EWD \cite{lu-etal-2024-entropy}, SWEET \cite{lee-etal-2024-wrote},  DIPmark  \cite{10.5555/3692070.3694260}, 
SIR \cite{DBLP:conf/iclr/LiuPHM024}, XSIR \cite{he-etal-2024-watermarks}, and sampling-based methods AAR \cite{aronsonpowerpoint}, SynthID \cite{dathathri2024scalable}. Detailed introduction and settings are provided in Appendix \ref{ap: baseline}.

\paragraph{Implementation Details}
We conduct experiments on two language models: OPT-1.3B \cite{zhang2022opt} and Llama-3.1-8B-Instruct \cite{dubey2024llama}. For more implementation details and hyperparameters, please refer to Appendix \ref{ap: imp}.

\subsection{Main Results}
Table \ref{tab: main result} presents the paraphrase attack robustness and spoofing attack robustness results on the RealNewsLike and BookSum datasets. Table \ref{tab: traceability} reports the spoofing attack traceability results on the RealToxicityPrompts and RTP-LX datasets. The watermark detectability  AUC for all methods is close to 1 and is provided in Figure \ref{fig: watermark detect}. We have the following observations and analyses:

\textbf{Current watermarking algorithms focus on paraphrase attacks but remain vulnerable to spoofing attacks.} While existing watermarking algorithms such as KGW exhibit strong and consistent robustness against paraphrase attacks, achieving an average AUC of approximately 0.9345, their performance drops markedly under spoofing attacks, where the average AUC falls to around 0.5042, revealing a significant vulnerability to malicious content manipulation. In comparison, DualGuard achieves an average AUC of \textbf{0.9410} across both paraphrase and spoofing attack settings, demonstrating an effective trade-off between defending against paraphrase and spoofing attacks.

\textbf{Content sensitivity property of the dual-stream watermark effectively detects spoofing attacks.}
Table~\ref{tab: main result} shows that baselines struggle to defend against spoofing attacks, with AUC values for spoofing attack robustness hovering around 0.5000, which severely limits reliable watermark detection. In contrast, DualGuard integrates content sensitivity into the watermark signal, allowing the model to leverage signal discrepancies between benign and malicious content to accurately detect spoofing attacks, achieving an average AUC of \textbf{0.9337}.

\begin{table*}[ht]
\centering
\resizebox{\linewidth}{!}{
    \begin{tabular}{lcccccccccccc}
    \toprule
     \multirow{2}{*}{\textbf{Method}} & \multicolumn{3}{c}{\textbf{\makecell{\textbf{Short Q, Short A}\\ \textbf{Factual Knowledge}}}} & \multicolumn{3}{c}{\textbf{\makecell{\textbf{Short Q, Long A}\\ \textbf{Long-form QA}}}} 
      & \multicolumn{3}{c}{\textbf{\makecell{\textbf{Long Q, Short A}\\ \textbf{Reasoning \& Coding}}}} & \multicolumn{3}{c}{\textbf{\makecell{\textbf{Long Q, Long A}\\ \textbf{Summarization}}}} \\
     \cmidrule(lr){2-4}  \cmidrule(lr){5-7} \cmidrule(lr){8-10} \cmidrule(lr){11-13}
     & \textbf{TPR} & \textbf{TNR}  & \textbf{GM} & \textbf{TPR} & \textbf{TNR}  & \textbf{GM} & \textbf{TPR} & \textbf{TNR}  & \textbf{GM} & \textbf{TPR} & \textbf{TNR} & \textbf{GM} \\
    \midrule
    \textbf{Original} & - & - & 33.00 & - & - & 23.87 & - & - & 51.19 & - & - & 22.03 \\
    \textbf{KGW} & 0.9450 & 0.6800 & 35.00 (+2.00)  & 0.9900 & 0.9800 & 23.86 (-0.01) & 0.8850 & 0.8000 & 48.84 (-2.35) & 0.9950 & 1.0000 & 21.66 (-0.37) \\
    \textbf{Unbiased} & 0.7800 & 0.6350 & 34.50 (+1.50) & 0.9600 & 0.9850 & 23.99 (+0.12) & 0.8100 & 0.6400 & 49.40 (-1.79) & 1.0000 &  0.9950 & 21.17 (-0.86) \\
    \textbf{AAR} & 1.0000 & 0.6950 & 37.50 (+4.50) & 0.9850 & 0.9800 & 23.19 (-0.68) & 0.7800 & 0.6950 & 59.82 (+8.63) & 0.7800 &  0.9800 & 21.35 (-0.68) \\
    \textbf{SynthID} & 0.9600 & 0.8600 & 25.50 (-7.50) & 0.9500 & 0.9650 & 23.93 (+0.06) & 0.9850 & 0.5450 & 50.85 (-0.34) & 1.0000 &  1.0000 & 20.71 (-1.32) \\
    \textbf{EWD} & 0.9600 & 0.7800 & 29.50 (-3.50) & 1.0000 & 1.0000 & 23.69 (-0.18) & 0.8350 & 0.9350 & 49.24 (-1.95) & 1.0000 &  1.0000 & 20.60 (-1.43) \\
    \textbf{SWEET} & 0.8250 & 0.8900 & 28.50 (-4.50) & 1.0000 & 1.0000 & 24.06 (+0.19) &  0.8250 & 0.8800 & 50.04 (-1.15) & 1.0000 &  1.0000 & 20.74 (-1.29) \\
    \textbf{DIPmark} & 0.8850 & 0.6750 & 33.00 (+0.00)  & 0.8850 & 0.9100 & 23.92 (+0.05) &  0.9350 & 0.5000 & 49.27 (-1.92) & 0.9900 &  0.9950 & 21.96 (-0.07) \\
    \textbf{SIR} & 0.9750 & 0.4750 & 35.00 (+2.00) & 1.0000 & 0.9250 & 22.34 (-1.53)  & 0.9700 & 0.7350 & 47.10 (-4.09) & 0.9800 & 0.9900 & 21.02 (-1.01) \\
    \textbf{XSIR} & 0.7750 & 0.9750 & 31.50 (-1.50) & 0.9000 & 0.8400 & 22.52 (-1.35) &  0.8100 & 0.6100 & 48.70 (-2.49) & 0.8800 &  0.9550 & 20.00 (-2.03) \\
    \textbf{DualGuard} & 0.9500 & 0.6500 & 35.50 (+2.50) & 0.9850 & 0.9850 & 23.35 (-0.52) & 0.9400 & 0.6550 & 49.88 (-1.31) & 1.0000 &  1.0000 & 19.64 (-2.39) \\
    \bottomrule
    \end{tabular}
}
\caption{Experimental results across different downstream tasks, evaluated using True Positive Rate (TPR), True Negative Rate (TNR), and Generation Metric (GM).}
\label{tab: downstream task}
\end{table*}

\textbf{Adaptive dual-stream watermarking effectively traces the spoofing attacks.} As shown in Table~\ref{tab: traceability}, existing methods struggle to trace spoofing attacks, with AUC values approaching 0.5000. In contrast, benefiting from the dual-stream watermarking design, our method enables continuous monitoring of adversarial watermark signals and accurately traces the spoofing attacks, achieving an average AUC of \textbf{0.8681}. These results demonstrate that our watermarking algorithm remains robust under complex real-world scenarios and provides an effective safeguard against LLM misuse.

\subsection{Impact of Attack Model}
We further investigate the robustness under different attack models, including GPT-4.1 (gpt-4.1-nano) \cite{achiam2023gpt}, Gemini-2.5 (gemini-2.5-flash-lite) \cite{team2024gemini}, and Qwen3 (qwen-flash) \cite{yang2025qwen3}. Figure \ref{fig: other attack} reports the results for Gemini-2.5 and Qwen3, while Figure \ref{fig: other attack gpt41} presents those for GPT-4.1, covering spoofing attack robustness on RealNewsLike and spoofing attack traceability on RealToxicityPrompts. DualGuard achieves consistently excellent performance in all scenarios, with average AUC scores of \textbf{0.9252} for spoofing attack robustness and \textbf{0.8898} for spoofing attack traceability. These results indicate that the proposed dual-stream watermark signal can accurately capture the characteristics of spoofing attacks, efficiently detect and trace spoofing attacks across diverse attack models, and thus achieve reliable and trustworthy watermark detection.

\begin{figure}[t]
\centering
\subfigure[Detectability, Robustness] {
    \includegraphics[width=0.45\linewidth]{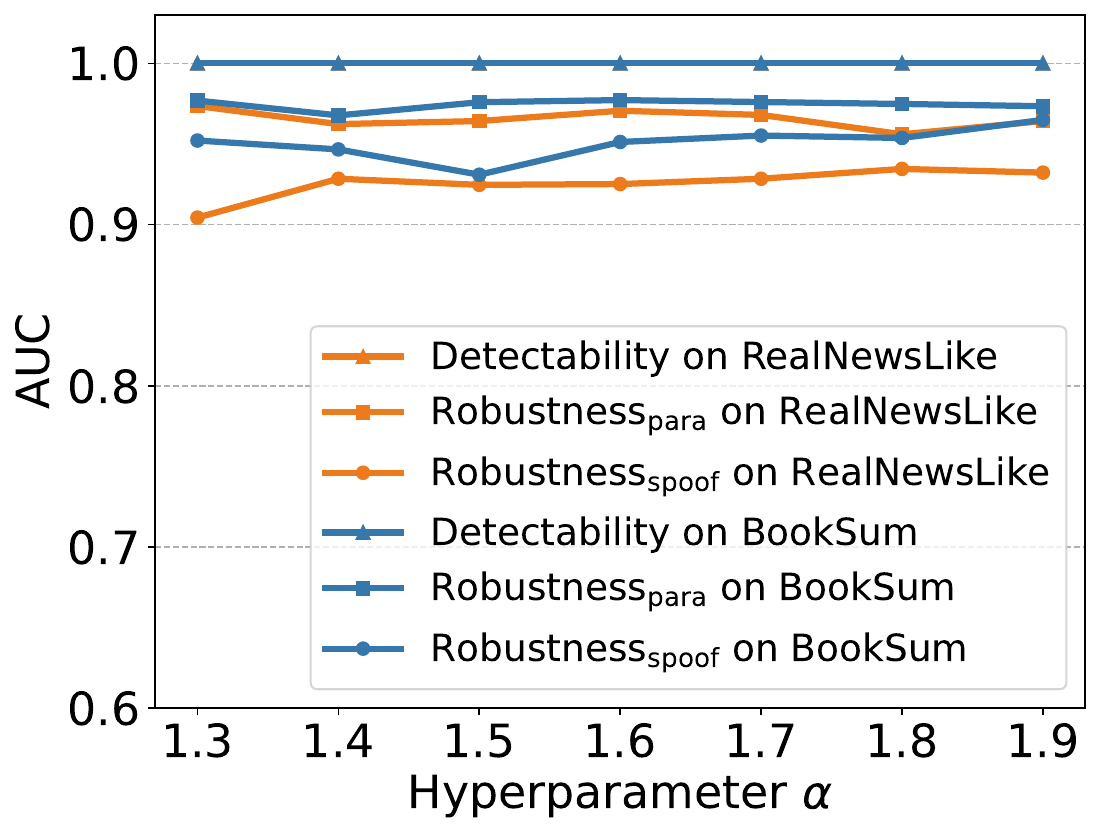}
}
\subfigure[Traceability] {
    \includegraphics[width=0.45\linewidth]{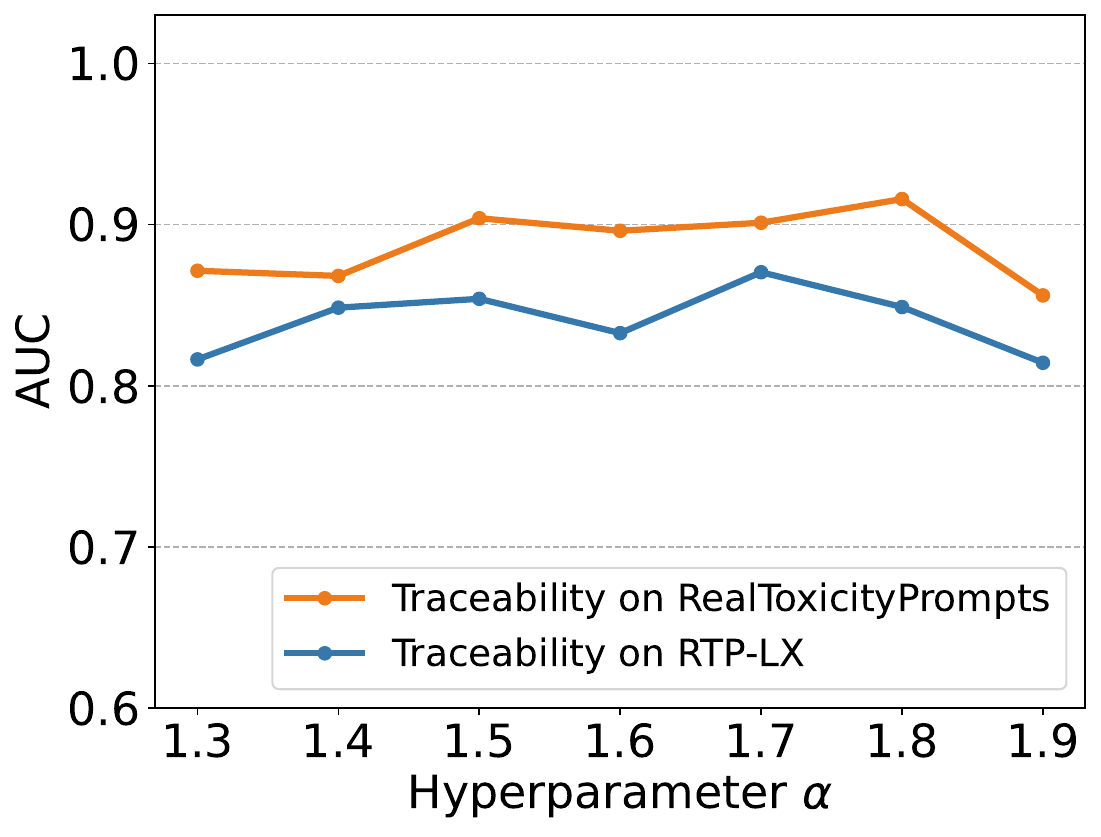}
}
\caption{The impact of Dual-stream selection.}
\label{fig: select}
\centering
\end{figure}

\subsection{Impact of Dual-stream Selection}
The parameter $\alpha$ serves as a critical threshold that determines which watermark head is selected during watermark injection. We analyze the impact of this key parameter on the overall performance in Figure \ref{fig: select}.  
Our method consistently demonstrates excellent performance across all parameter settings. On the RealNewsLike and BookSum datasets, the average AUC for watermark detectability reaches \textbf{1.0000} and \textbf{1.0000}, the average AUC for paraphrase attack robustness achieves \textbf{0.9655} and \textbf{0.9746}, and the average AUC for spoofing attack robustness attains \textbf{0.9254} and \textbf{0.9507}.
On the RealToxicityPrompts and RTP-LX datasets, the average AUC for spoofing attack traceability is \textbf{0.8875} and \textbf{0.8407}. Considering the overall performance, selecting the moderate threshold for $\alpha$ achieves the more stable and balanced performance (e.g., 1.7). These results demonstrates that our method is largely insensitive to hyperparameter variations, ensuring stable and generalizable performance across diverse scenarios.

\subsection{Text Quality Analysis}
We evaluate the impact of watermarking algorithms on text quality by measuring downstream task performance and perplexity. Downstream task performance on WaterBench \cite{tu-etal-2024-waterbench} is reported in Table~\ref{tab: downstream task}, and perplexity results are shown in Figure~\ref{fig: ppl}. Detailed settings are provided in Appendix~\ref{ap: quality}. Compared to the baseline without watermark injection (Original) and other watermarking methods, DualGuard introduces only a minor impact on text quality. Specifically, DualGuard achieves competitive perplexity scores, with an average decrease of approximately \textbf{0.43}\% in the generation metrics across the four downstream tasks. These results indicate that DualGuard preserves text quality while maintaining reliable watermark detectability, making it suitable for real-world deployment.

\section{Conclusion}
In this paper, we comprehensively investigate the capability of watermarking algorithms to defend against paraphrase and spoofing attacks. Defending against spoofing attacks remains a pressing and underexplored challenge in existing watermarking research. To address this issue, we propose DualGuard, an adaptive dual-stream watermarking algorithm that represents the first watermarking framework capable of resisting both paraphrase and spoofing attacks.
The design of dual-stream watermarking enables our scheme to not only enhance the robustness against paraphrase and spoofing attacks, but also accurately track spoofing attacks.

\section*{Acknowledgments}
This work is supported by the Postdoctoral Fellowship Program of CPSF under Grant Number GZC20251076, and the National Natural Science Foundation of China (No.U2336202).

\section*{Limitations}
This paper investigates the vulnerability of current watermarking algorithms to piggyback spoofing attacks, showing that misattributing generated content to watermark deployers can severely undermine watermark reliability. DualGuard incorporates the watermark head selection into the watermark mapping model, which can be combined with LLM's content detection strategy in future work to improve the overall system's efficiency and security. Furthermore, the dual-stream watermarking mechanism can be extended to other domains, including vision tasks. We believe that the dual-stream watermarking concept provides a new perspective and direction for watermarking schemes, advancing their reliability and practical applicability.

\section*{Ethical Statement}
With the rapid advancement of large language models, users can easily generate high-quality, human-like content, which has simultaneously raised growing concerns about their potential misuse. As a promising countermeasure, language model watermarking aims to actively embed identifiable signals into generated text to enable the reliable detection of machine-generated content. However, in real-world scenarios, watermarking methods remain vulnerable to various attacks, such as paraphrase and piggyback spoofing attacks, which substantially undermine their reliability and security. In this paper, we propose a simple yet effective method to enhance the robustness of watermarking algorithms against both paraphrase and piggyback spoofing attacks. By strengthening the security guarantees of watermarking under these realistic adversarial settings, our method takes an important step toward more trustworthy and deployable watermarking systems. We hope that this work will motivate future research interests in the reliability and security of language model watermarks.

\bibliography{custom}

\cleardoublepage

\appendix

\section{Appendix Structure}
The appendix is organized into three main parts:
\paragraph{Baselines, datasets, and experimental settings} 
Appendix \ref{ap: related} presents additional related work. Appendix \ref{ap: dataset} describes the data analysis and processing procedures. Appendix \ref{ap: metrics} introduces the evaluation metrics. Appendix \ref{ap: baseline} presents the baseline settings. Appendix \ref{ap: imp} provides implementation details. Appendix \ref{ap: prompt} presents prompts used for paraphrase and piggyback spoofing attacks.
\paragraph{Ablation studies and parameter analysis}
Appendix \ref{ap: ablation} presents the ablation study. Appendix \ref{ap: prefix} analyzes the impact of the watermark prefix length. Appendix \ref{ap: window} investigates the impact of window length. Appendix \ref{ap: token} studies the impact of token length. Appendix \ref{ap: encoding} explores the impact of encoding models. Appendix \ref{ap: spoof analysis} analyzes the effectiveness of spoofing attacks. Appendix \ref{ap: detector} explores the impact of detector selection. Appendix \ref{ap: decoupled} analyzes the performance of decoupled baselines.

\paragraph{Generalization and robustness analysis}
Appendix \ref{ap: semantic-preserving} evaluates the robustness against various semantic-preserving attacks. Appendix \ref{ap: quality} evaluates the impact of watermarking algorithms on text quality. Appendix \ref{ap: multilingual} evaluates the generalization of DualGuard in multilingual scenarios. Appendix \ref{ap: diff detector} explores the impact of different detectors on performance. 
Appendix \ref{ap: diff model} evaluates the performance of DualGuard across different models. Appendix \ref{ap: temperature} analyzes the impact of generation temperature.  Appendix \ref{ap: security} analyzes the security of DualGuard. Appendix \ref{ap: complexity} presents the complexity analysis.

\section{More Related Works}
\label{ap: related}
\subsection{Language Model Watermarking}
Logits-based methods embed watermark signals by directly modifying the output logits of the language model. KGW \cite{kirchenbauer2023watermark} is a representative logits-based method that randomly partitions the LLM vocabulary into green and red lists, and then increases the logits of tokens in the green list, thereby encouraging the watermarked text to contain a higher proportion of green-list tokens. 
Unbiased \cite{DBLP:conf/iclr/HuCWWZH24} and DIPmark \cite{10.5555/3692070.3694260} further introduce unbiased watermarking techniques that ensure identical expected distributions between watermarked and un-watermarked texts, thereby preserving the original token probability distribution. Furthermore, SIR \cite{DBLP:conf/iclr/LiuPHM024}, XSIR \cite{he-etal-2024-watermarks}, and Adaptive \cite{10.5555/3692070.3693308} leverage semantic embeddings to derive watermark logits, while EWD \cite{lu-etal-2024-entropy} and SWEET \cite{lee-etal-2024-wrote} inject watermarks from the entropy-based perspective. In contrast, sampling-based methods embed the watermark message by guiding the token sampling process. AAR \cite{aronsonpowerpoint} employs exponential minimum sampling to embed watermarks, while SynthID \cite{dathathri2024scalable} introduces the tournament-based sampling scheme that preserves text quality while ensuring watermark detectability. Furthermore, SemStamp \cite{hou-etal-2024-semstamp} and k-SemStamp \cite{hou-etal-2024-k} propose sentence-level sampling algorithms, which leverage locality-sensitive hashing \cite{10.1145/276698.276876} and $k$-means clustering \cite{lloyd1982least} to partition the semantic space into watermarked and non-watermarked regions, ensuring that the generated sentences originate from the watermarked region.

\begin{algorithm}[t]
\small
\caption{Watermarked Text Generation}
\begin{algorithmic}[1]
    \STATE \textbf{Input:} LLM $\mathcal{M}$, encoding model $\mathcal{E}$, watermark head $\Theta_s$ and $\Theta_a$, watermark prefix length $\rho$, window length $k$, scaling  parameter $\delta$
    \STATE \textbf{Output:} Generated text $y$

    \STATE Initialize watermark head  $\Theta \leftarrow \Theta_s$

    \FOR{$t= 1$ to $T$}
    \STATE // Select watermark head
    \IF{$t \bmod k = 1$}
    \STATE $\Theta \leftarrow 
    \begin{cases}
        {\Theta}_s,  \operatorname{dist}(y_{:t}) < \alpha \\
        {\Theta}_a,  \operatorname{dist}(y_{:t}) \geq \alpha 
    \end{cases}$
    \ENDIF
    \STATE Generate the next token logits $P_\mathcal{M}^t \leftarrow \mathcal{M}(y_{:t})$ 
    \STATE Generate current embedding $e_t \leftarrow \mathcal{E}(y_{t-\rho:t})$
    \STATE Generate watermark logits $P_\Theta^t \leftarrow \Theta(e_t)$ \hfill 
    \STATE Insert watermark $P_{\mathcal{M}'}^t \leftarrow P_{\mathcal{M}}^t + \delta \cdot P_{\mathcal{M}}^t P_\Theta^t$ 
    \STATE Generate the next token $y_t \leftarrow P_{\mathcal{M}'}^t$
    \ENDFOR    
\end{algorithmic}
\label{algo: gen}
\end{algorithm}

\subsection{Watermark Robustness}
Robustness against watermark removal attacks is a key metric for watermarking algorithms, since text can be easily modified (e.g., paraphrased). Recent algorithms have enhanced robustness in several ways. For instance, SWEET \cite{lee-etal-2024-wrote} embeds watermarks in high-entropy segments based on an entropy threshold, while EWD \cite{lu-etal-2024-entropy} assigns higher influence weights to high-entropy tokens during watermark detection. In addition, SIR \cite{DBLP:conf/iclr/LiuPHM024}, XSIR \cite{he-etal-2024-watermarks}, and Adaptive \cite{10.5555/3692070.3693308} train watermark models to generate the semantic invariant watermark, and \citet{lau-etal-2024-waterfall} employs LLM-based paraphrasing to embed watermarks while preserving semantic content.
However, robustness against spoofing attacks remains largely underexplored \cite{DBLP:conf/nips/Pang0ZS24}. More critically, focusing solely on robustness against watermark removal attacks introduces a vulnerability that enables adversaries to perform piggyback spoofing attacks. In such cases, watermarking algorithms may still detect the watermark in altered text, thereby creating an opportunity for adversaries to inject malicious content while retaining a valid watermark signal.

\begin{algorithm}[t]
\small
\caption{Watermark Detection}
\begin{algorithmic}[1]
    \STATE \textbf{Input:} Text $y$, LLM $\mathcal{M}$, encoding model $\mathcal{E}$, watermark head $\Theta_s$ and $\Theta_a$, watermark prefix length $\rho$, window length $k$, threshold $\theta_{\text{wd}}$, $\theta_{\text{sd}}$, $\theta_{\text{st}}$
    \STATE \textbf{Output:} Text $y$ label
    \STATE Initialize score $\text{Score}_{\text{wd}}$, $\text{Score}_{\text{sd}}$, $\text{Score}_{\text{st}}$
   \STATE Initialize watermark head  $\Theta \leftarrow \Theta_s$
    
    \FOR{$t= 1$ to $T$}
    \STATE Generate current text embedding $e_t \leftarrow \mathcal{E}(y_{t-\rho:t})$
    % \STATE // Select watermark head
    \IF{$t \bmod k = 1$}
    \STATE $\Theta \leftarrow 
    \begin{cases}
        {\Theta}_s,  \operatorname{dist}(y_{:t}) < \alpha \\
        {\Theta}_a,  \operatorname{dist}(y_{:t}) \geq \alpha 
    \end{cases}$ 
    
    \STATE Calculate spoofing attack detection score 
    \STATE $\text{Score}_{\text{sd}} \leftarrow  \operatorname{dist}(y_{:t})$ 

    \ENDIF
    \STATE Generate watermark logits $P_\Theta^t \leftarrow \Theta(e_t)$
    \STATE Calculate watermark score $\text{Score}_{\text{wd}} \leftarrow P_\Theta^t[y_t]$
    \STATE Calculate spoofing attack tracing score
    \IF{$\Theta$ is $\Theta_a$}
       \STATE $\text{Score}_{\text{st}} \leftarrow P_\Theta^t[y_t]$
    \ENDIF
    \ENDFOR

    \IF{$\text{Score}_{\text{wd}} < \theta_{\text{wd}}$}
        \STATE \textbf{Return:} un-watermarked text
    \ELSE
        \IF{$\text{Score}_{\text{sd}} < \theta_{\text{sd}}$}
        \STATE \textbf{Return:} watermarked text
        \ELSE
        \IF{$\text{Score}_{\text{st}} < \theta_{\text{st}}$}
        \STATE \textbf{Return:} spoofed watermarked text
        \ELSE
        \STATE \textbf{Return:} malicious watermarked text
        \ENDIF
        \ENDIF
    \ENDIF    
\end{algorithmic}
\label{algo: detect}
\end{algorithm}

\section{Dataset}
\label{ap: dataset}
\paragraph{Colossal Clean Crawled Corpus} (C4) \cite{10.5555/3455716.3455856} is a large-scale corpus constructed from the public Common Crawl web archive through extensive cleaning and filtering. Its RealNewsLike subset further restricts the corpus to documents originating from domains associated with the RealNews dataset, thereby retaining content that primarily reflects the news domain.

\paragraph{BookSum} \cite{kryscinski-etal-2022-booksum} is a long-form narrative summarization dataset drawn from the literature domain, covering diverse narrative forms such as novels, plays, and stories. It provides highly abstractive and human-written summaries at three hierarchical levels of increasing difficulty: paragraph, chapter, and book, supporting research on multi-level and long-context summarization.

\begin{figure}[t]
\centering
\includegraphics[width=1.0\linewidth]{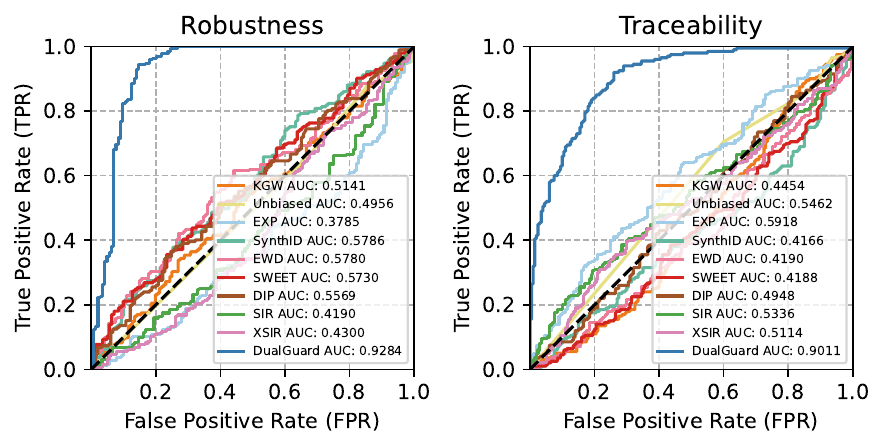}
\caption{Experimental results of GPT-4.1 on RealNewsLike and RealToxicityPrompts dataset.}
\label{fig: other attack gpt41}
\centering
\end{figure}

\paragraph{RealToxicityPrompts} \cite{gehman-etal-2020-realtoxicityprompts} is a large-scale prompt dataset constructed from a broad corpus of English web text. It comprises a diverse collection of benign and toxic sentence-level prompts, designed to evaluate the tendency of large language models to produce toxic or harmful content during text generation.

\paragraph{RTP-LX} \cite{de2025rtp} is a multilingual corpus comprising human-translated and human-annotated malicious prompts across multiple languages. It is designed to evaluate the capability of LLMs to generate toxic content in culturally sensitive and multilingual scenarios.

\begin{figure*}[h]
\centering
\subfigure[OPT-1.3B]{
\begin{minipage}[t]{0.5\linewidth}
\centering
\includegraphics[width=\linewidth]{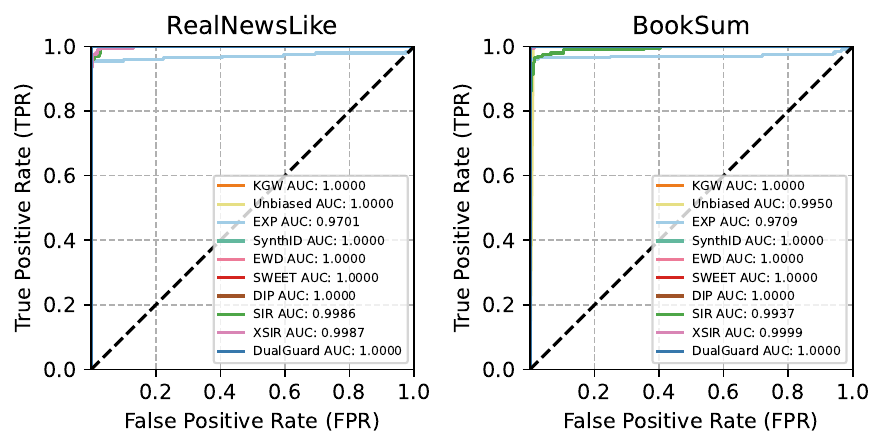}
\end{minipage}%
}%
\subfigure[Llama3.1-8B-Instruct]{
\begin{minipage}[t]{0.5\linewidth}
\centering
\includegraphics[width=\linewidth]{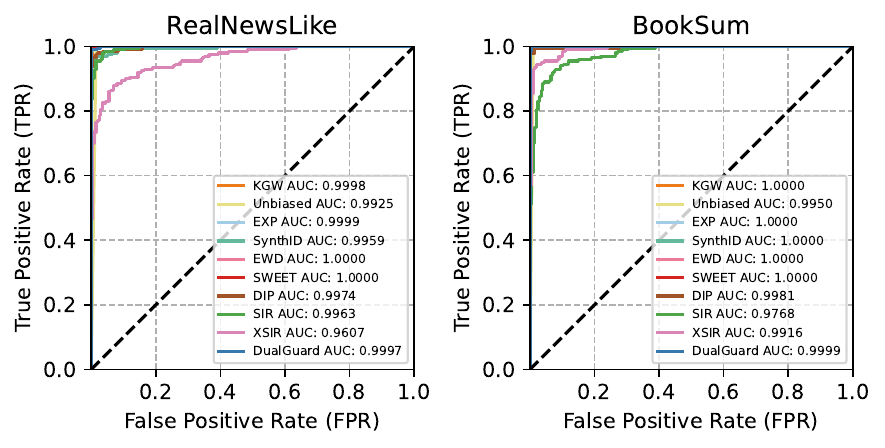}
\end{minipage}%
}%
\caption{Experimental results of watermark detectability on RealNewsLike and BookSum datasets.}
\label{fig: watermark detect}
\end{figure*}

For each dataset, we follow the official data split. 
Following \citet{he-etal-2024-watermarks}, we randomly sample 200 instances from the RealNewsLike and BookSum datasets. For each instance, the first 20 tokens are used as prompts, while the last 200 tokens are treated as natural human-written text.  Conditioned on these prompts, we generate $T = 200 \pm 30$ tokens using the watermarking algorithm to obtain watermarked text.
Additionally, we randomly sample 500 benign and malicious prompts from the RealToxicityPrompts and RTP-LX datasets. In the RealToxicityPrompts dataset, benign and malicious prompts are distinguished according to their severe toxicity score labels. 
For the RTP-LX dataset, we use its English subset containing only malicious prompts and construct benign prompts by truncating benign completion texts to match the length of the malicious ones. To evaluate spoofing attacks, we employ a binary classification detector (SiEBERT \cite{hartmann2023}) to filter the generated texts and retain up to 200 validated instances, ensuring that each spoofed text is successfully identified as malicious, meaning that the spoofing attack effectively produces harmful content.

\section{Evaluation Metrics}
\label{ap: metrics}
We employ the following metrics to evaluate watermarking algorithms:
\paragraph{Watermark Detectability} which measures whether watermarked text can be reliably detected, is a fundamental requirement for watermarking algorithms. In this setting, watermarked text is treated as the positive sample, while human-written text serves as the negative sample.
\paragraph{Paraphrase Attack Robustness} which measures whether watermarked text that has been subjected to paraphrase attacks can be reliably detected. In this setting, paraphrased watermarked text is treated as the positive sample, while human-written text serves as the negative sample.
\paragraph{Spoofing Attack Robustness} which measures whether malicious content embedded through spoofing attacks in watermarked text can be reliably detected. In this setting, watermarked text that has been subjected to spoofing attacks is treated as the positive sample, and paraphrased watermarked text serves as the negative sample.\footnote{Paraphrasing constitutes the prerequisite of spoofing attacks, thus providing the most challenging hard negative for distinguishing adversarially injected malicious content.}
\paragraph{Spoofing Attack Traceability} which measures the ability to attribute the source of malicious content in watermarked text. In this setting, malicious watermarked text generated through spoofing attacks is treated as the positive sample, while malicious watermarked text generated directly by LLMs and then paraphrased serves as the negative sample. 

All metrics are evaluated using the area under the receiver operating characteristic curve (\textbf{AUC}) and the true positive rate when the false positive rate is 5\% or 10\% (\textbf{TP@5\%}, \textbf{TP@10\%}).

\begin{table*}[ht]
\centering
\resizebox{\linewidth}{!}{
    \begin{tabular}{lcccccccccccc}
    \toprule
    \multirow[c]{3}{*}{\textbf{Method}} & \multicolumn{9}{c}{\textbf{RealNewsLike}} & \multicolumn{3}{c}{\textbf{RealToxicityPrompts}} \\
    \cmidrule(lr){2-10}  \cmidrule(lr){11-13}
    & \multicolumn{3}{c}{\textbf{Detectability}}
    & \multicolumn{3}{c}{$\textbf{Robustness}_{\textbf{para}}$} &  \multicolumn{3}{c}{$\textbf{Robustness}_{\textbf{spoof}}$} & 
    \multicolumn{3}{c}{$\textbf{Traceability}_{\textbf{spoof}}$} \\
    \cmidrule(lr){2-4}  \cmidrule(lr){5-7}     \cmidrule(lr){8-10}  \cmidrule(lr){11-13}
    & \textbf{AUC} & \textbf{TP@5\%} & \textbf{TP@10\%} & \textbf{AUC} & \textbf{TP@5\%} & \textbf{TP@10\%} & \textbf{AUC} & \textbf{TP@5\%} & \textbf{TP@10\%} & \textbf{AUC} & \textbf{TP@5\%} & \textbf{TP@10\%}\\
    \midrule
    \textbf{DualGuard} & 1.0000 & 1.0000  & 1.0000 & 0.9680 & 0.8600  & 0.9250 & 0.9284 & 0.3505  & 0.8247 & 0.9011 & 0.5200  & 0.6600  \\
    \textbf{-w/o $\mathcal{L}_{\text{con}}$} & 1.0000 & 1.0000 & 1.0000 & 0.9732 & 0.8800 & 0.9050 & 0.4554 & 0.0153 & 0.0306 & 0.5000 & 0.0500 &  0.1000  \\
    \midrule
    & \multicolumn{9}{c}{\textbf{BookSum}} & \multicolumn{3}{c}{\textbf{RTP-LX }} \\
    \cmidrule(lr){1-10}  \cmidrule(lr){11-13}
    \textbf{DualGuard} & 1.0000 & 1.0000 & 1.0000 & 0.9760 & 0.9200  & 0.9550 & 0.9552 & 0.7784  & 0.8693 & 0.8704 &  0.5100  & 0.6500  \\
    \textbf{-w/o $\mathcal{L}_{\text{con}}$}  & 1.0000 & 1.0000 & 1.0000 & 0.9919 & 0.9700 & 0.9800 & 0.4770 & 0.0330 & 0.0824 & 0.5000 & 0.0500 & 0.1000  \\
    \bottomrule
    \end{tabular}
    }
\caption{Ablation study of the content sensitivity loss on four datasets using the OPT-1.3B model.}
\label{tab: ablation}
\end{table*}

\section{Baseline}
\label{ap: baseline}
We evaluate DualGuard against following methods:
\begin{itemize}
    \item KGW \cite{kirchenbauer2023watermark}: which splits the LLM’s vocabulary into green and red lists, and injects watermarks by enhancing the probability of green tokens. The parameters are set as follows: $\gamma$ = 0.5, $\delta$ = 2.0, hash\_key = 15485863, prefix\_length = 1, window\_scheme = left.
    \item Unbiased \cite{DBLP:conf/iclr/HuCWWZH24}: which proposes $\delta$-reweight and $\gamma$-reweight watermarking techniques, which can integrate watermarks without affecting the output probability distribution with appropriate implementation. The parameters are set as follows: $\gamma$ = 0.5, key = 42, prefix\_length = 5.
    \item AAR \cite{aronsonpowerpoint}: which employs exponential minimum sampling to embed watermarks. The parameters are set as follows: prefix\_length = 4, hash\_key = 15485863, sequence\_length = 200.
    \item SynthID \cite{dathathri2024scalable}: which introduces the tournament-based sampling scheme to inject watermarks. The parameters are set as follows: ngram\_len = 5, sampling\_size = 65536, seed = 0, mode = ``non-distortionary'', num\_leaves =2, context\_size = 1024, detector\_type = ``mean''.
    \item EWD \cite{lu-etal-2024-entropy}: which is an entropy-based watermarking algorithm that gives higher-entropy tokens higher influence weights during watermark detection. The parameters are set as follows: $\gamma$ = 0.5, $\delta$ = 2.0,hash\_key = 15485863, prefix\_length = 1.
    \item SWEET \cite{lee-etal-2024-wrote}: which proposes a selective watermarking method based on entropy threshold applied to code generation scenarios, which achieves significantly higher performance in machine-generated code detection while maintaining code quality. The parameters are set as follows: $\gamma$ = 0.5, $\delta$ = 2.0, hash\_key = 15485863, entropy\_threshold = 0.9, prefix\_length = 1.
    \item DIPmark  \cite{10.5555/3692070.3694260}: which preserves the original token distribution during watermark generation and modifies the token distribution through a reweight function to enhance the probability of these selected tokens during sampling. The parameters are set as follows: $\gamma$ = 0.5, $\alpha$ = 0.45, key = 42, hash\_key = 15485863, prefix\_length = 5.
    \item SIR \cite{DBLP:conf/iclr/LiuPHM024}: which is a semantic invariant robust watermark that transforms semantic embeddings into watermark logits through a watermark model. The parameters are set as follows: $\delta$ = 1.0, chunk\_length = 10, scale\_dimension = 300.
    \item XSIR \cite{he-etal-2024-watermarks}: which applies SIR to the cross-lingual scenario by clustering semantically equivalent tokens and applying a multilingual watermark model. The parameters are set as follows: $\delta$ = 1.0, chunk\_length = 10, scale\_dimension = 300.

\end{itemize}

\section{Implementation Details}
\label{ap: imp}
We conduct experiments on two language models: OPT-1.3B \cite{zhang2022opt} and Llama-3.1-8B-Instruct \cite{dubey2024llama}. We employ OPT-1.3B as the foundational language model in our experiments. The encoding model $\mathcal{E}$ used is C-BERT \cite{chanchani-huang-2023-composition}. 

In the watermark mapping model training stage, the watermark mapping model is trained on the STSBenchmark \cite{muennighoff2022mteb}, a semantic textual similarity dataset, employing SiEBERT \cite{hartmann2023} as the binary classification detector to divide text into benign and malicious subsets, with the detection threshold set to 0.5.
The model is optimized using the AdamW optimizer with the learning rate of $2 \times 10^{-4}$. The scale factor $\tau$ in the semantic loss is set to 20, the margin parameter $\eta$ in the contrastive loss is set to 0.9, and the corresponding weight $\lambda$ is set to 1.0. The watermark mapping model is model-agnostic and can be applied across different LLMs.

In the watermark injection stage,  the threshold $\alpha$ for selecting the watermark head is set to 1.7, the watermark prefix length $\rho$ is set to 1, and the fixed-length window length $k$ is set to 12. The scale factor $\gamma$ for watermark logits generation is $1 \times 10^{3}$, while the watermark weights $\delta$ are set to 0.1 and 0.06 for OPT-1.3B and Llama-3.1-8B-Instruct. 

All models and datasets are accessible via HuggingFace, and all experiments are conducted on NVIDIA A100 80GB GPUs.
Baseline implementations follow the official repositories and the MarkLLM toolkit \cite{pan-etal-2024-markllm}.

\section{Prompt}
\label{ap: prompt}
Following \citet{an2025defending}, the prompts used for paraphrase and piggyback spoofing attacks are shown in Figure~\ref{fig: paraphrase prompt} and Figure~\ref{fig: spoof prompt}, respectively.
Paraphrase attacks rewrite the original text while preserving its semantics, whereas piggyback spoofing attacks go a step further by injecting malicious or harmful content into the text.

\begin{table*}[ht]
\centering
\resizebox{\linewidth}{!}{
    \begin{tabular}{ccccccccccccc}
    \toprule
    \multirow[c]{3}{*}{\textbf{Length}} & \multicolumn{9}{c}{\textbf{RealNewsLike}} & \multicolumn{3}{c}{\textbf{RealToxicityPrompts }} \\
    \cmidrule(lr){2-10}  \cmidrule(lr){11-13}
    & \multicolumn{3}{c}{\textbf{Detectability}}
    & \multicolumn{3}{c}{$\textbf{Robustness}_{\textbf{para}}$} &  \multicolumn{3}{c}{$\textbf{Robustness}_{\textbf{spoof}}$} & 
    \multicolumn{3}{c}{$\textbf{Traceability}_{\textbf{spoof}}$} \\
    \cmidrule(lr){2-4}  \cmidrule(lr){5-7}     \cmidrule(lr){8-10}  \cmidrule(lr){11-13}
    & \textbf{AUC} & \textbf{TP@5\%} & \textbf{TP@10\%} & \textbf{AUC} & \textbf{TP@5\%} & \textbf{TP@10\%} & \textbf{AUC} & \textbf{TP@5\%} & \textbf{TP@10\%} & \textbf{AUC} & \textbf{TP@5\%} & \textbf{TP@10\%}\\
    \midrule
    \textbf{1} & 1.0000 & 1.0000  & 1.0000 & 0.9680  & 0.8600  & 0.9250 & 0.9284 & 0.3505  & 0.8247 & 0.9011 & 0.5200  & 0.6600   \\
    \textbf{2} & 1.0000 & 1.0000  & 1.0000 & 0.9590 & 0.7150  & 0.9150 & 0.9283 & 0.4896  & 0.6979 & 0.9132 & 0.5150  & 0.7400  \\
    \textbf{3} & 1.0000 & 1.0000  & 1.0000 & 0.9362 & 0.7050  & 0.7800 & 0.8980 & 0.3069  & 0.5661 & 0.9100 & 0.5400  & 0.7350  \\
    \textbf{4} & 1.0000 & 1.0000  & 1.0000 & 0.9147 & 0.5450  & 0.7650 &  0.9298 & 0.5648  & 0.7565 & 0.9061 & 0.3600 & 0.6750  \\

    \bottomrule
    \end{tabular}
    }
\caption{The impact of watermark prefix length $\rho$ on watermark detectability, paraphrase attack robustness, and spoofing attack robustness and traceability using the OPT-1.3B model.}
\label{tab: prefix length}
\end{table*}

\begin{table*}[ht]
\centering
\resizebox{\linewidth}{!}{
    \begin{tabular}{ccccccccccccc}
    \toprule
    \multirow[c]{3}{*}{\textbf{Length}} & \multicolumn{9}{c}{\textbf{RealNewsLike}} & \multicolumn{3}{c}{\textbf{RealToxicityPrompts }} \\
    \cmidrule(lr){2-10}  \cmidrule(lr){11-13}
    & \multicolumn{3}{c}{\textbf{Detectability}}
    & \multicolumn{3}{c}{$\textbf{Robustness}_{\textbf{para}}$} &  \multicolumn{3}{c}{$\textbf{Robustness}_{\textbf{spoof}}$} & 
    \multicolumn{3}{c}{$\textbf{Traceability}_{\textbf{spoof}}$} \\
    \cmidrule(lr){2-4}  \cmidrule(lr){5-7}     \cmidrule(lr){8-10}  \cmidrule(lr){11-13}
    & \textbf{AUC} & \textbf{TP@5\%} & \textbf{TP@10\%} & \textbf{AUC} & \textbf{TP@5\%} & \textbf{TP@10\%} & \textbf{AUC} & \textbf{TP@5\%} & \textbf{TP@10\%} & \textbf{AUC} & \textbf{TP@5\%} & \textbf{TP@10\%}\\
    \midrule
    \textbf{4} & 1.0000 & 1.0000  & 1.0000 & 0.9714 & 0.8250  & 0.9000 & 0.9226 & 0.5104  & 0.7344 & 0.9037 & 0.5200  & 0.7300  \\
    \textbf{6} & 1.0000 & 1.0000  & 1.0000 & 0.9754 & 0.8750  & 0.9250 & 0.9077 & 0.4479  & 0.6458 & 0.8747 & 0.3500  & 0.5550  \\
    \textbf{8} & 0.9990 & 0.9950  & 0.9950 & 0.9645 & 0.8850  & 0.9200 & 0.9559 & 0.7000  & 0.8895 & 0.8956 & 0.5150  & 0.6000  \\
    \textbf{10} & 1.0000 & 1.0000  & 1.0000 & 0.9637 & 0.8050  & 0.9000 & 0.9410 & 0.4688  & 0.8750 & 0.9150 & 0.5050  & 0.7300  \\
    \textbf{12} & 1.0000 & 1.0000  & 1.0000 & 0.9680  & 0.8600  & 0.9250 & 0.9284 & 0.3505  & 0.8247 & 0.9011 & 0.5200  & 0.6600  \\

    \bottomrule
    \end{tabular}
    }
\caption{The impact of window length $k$ on watermark detectability, paraphrase attack robustness, and spoofing attack robustness and traceability using the OPT-1.3B model.}
\label{tab: window length}
\end{table*}

\section{Ablation Study}
\label{ap: ablation}
The content sensitivity property is critical for enabling DualGuard to defend against piggyback spoofing attacks. We conduct ablation studies to investigate the effectiveness of the content sensitivity property, and the results are shown in Table \ref{tab: ablation}. Specifically, w/o $\mathcal{L}_{\text{con}}$ denotes training the watermark mapping model $\mathcal{G}$ without the content sensitivity loss $\mathcal{L}_{\text{con}}$. Removing $\mathcal{L}_{\text{con}}$ slightly improves robustness against paraphrasing attacks, since the model no longer needs to select between different watermark heads. However, this variant completely fails to defend against piggyback spoofing attacks, achieving average AUC scores of only 0.4662 and 0.5000 for spoofing attack robustness and traceability. These results indicate that, without content sensitivity, the watermark signal cannot reliably distinguish benign semantic variations from malicious content manipulation. In contrast, DualGuard explicitly incorporates dual-stream watermarking guided by content sensitivity, allowing the model to dynamically activate different watermark signals for benign and malicious content. As a result, DualGuard simultaneously preserves robustness against paraphrasing attacks while effectively detecting and tracing piggyback spoofing attacks, leading to reliable and verifiable watermark detection.

\section{Impact of Watermark Prefix Length}
\label{ap: prefix}
In this section, we analyze the impact of the watermark prefix length $\rho$ on watermark detectability, robustness to paraphrasing attacks, and robustness and traceability under spoofing attacks. The results are summarized in Table~\ref{tab: prefix length}, where $\rho \in \{1, 2, 3, 4\}$. Across all prefix length settings, DualGuard consistently achieves an average AUC of \textbf{1.0000} for watermark detectability, along with AUC scores of \textbf{0.9445} for paraphrasing robustness and \textbf{0.9211} and \textbf{0.9076} for spoofing robustness and traceability, respectively. These results demonstrate that DualGuard consistently preserves strong watermark detectability while remaining robust to both paraphrasing and spoofing attacks under varying watermark prefix lengths. Overall, the observed performance stability with respect to $\rho$ suggests that DualGuard is largely insensitive to the choice of watermark prefix length, making it suitable for practical deployment scenarios in which different prefix configurations may be required.

\begin{table*}[ht]
\centering
\resizebox{\linewidth}{!}{
    \begin{tabular}{lcccccccccccc}
    \toprule
    \multirow[c]{3}{*}{\textbf{Model}} & \multicolumn{9}{c}{\textbf{RealNewsLike}} & \multicolumn{3}{c}{\textbf{RealToxicityPrompts }} \\
    \cmidrule(lr){2-10}  \cmidrule(lr){11-13}
    & \multicolumn{3}{c}{\textbf{Detectability}}
    & \multicolumn{3}{c}{$\textbf{Robustness}_{\textbf{para}}$} &  \multicolumn{3}{c}{$\textbf{Robustness}_{\textbf{spoof}}$} & 
    \multicolumn{3}{c}{$\textbf{Traceability}_{\textbf{spoof}}$} \\
    \cmidrule(lr){2-4}  \cmidrule(lr){5-7}     \cmidrule(lr){8-10}  \cmidrule(lr){11-13}
    & \textbf{AUC} & \textbf{TP@5\%} & \textbf{TP@10\%} & \textbf{AUC} & \textbf{TP@5\%} & \textbf{TP@10\%} & \textbf{AUC} & \textbf{TP@5\%} & \textbf{TP@10\%} & \textbf{AUC} & \textbf{TP@5\%} & \textbf{TP@10\%}\\
    \midrule
    \rowcolor[gray]{0.9} \multicolumn{13}{c}{OPT-1.3B} \\
    \textbf{C-BERT} & 1.0000 & 1.0000  & 1.0000 & 0.9680 & 0.8600  & 0.9250 & 0.9284 & 0.3505  & 0.8247 & 0.9011 & 0.5200  & 0.6600  \\
    \textbf{S-BERT} & 1.0000 & 1.0000  & 1.0000 & 0.9351 & 0.7500  & 0.8200 & 0.9376 & 0.6406  & 0.7760 & 0.8796 & 0.4850  & 0.5650  \\
    \midrule
    \rowcolor[gray]{0.9} \multicolumn{13}{c}{Llama3.1-8B-Instruct} \\
    \textbf{C-BERT} & 0.9997 & 1.0000  & 1.0000 & 0.9244 & 0.6200  & 0.7600 & 0.9159 & 0.2552  & 0.6562 & 0.8513 & 0.3750  & 0.5700  \\
    \textbf{S-BERT} & 1.0000 & 1.0000  & 1.0000 & 0.9568 & 0.8150  & 0.9200 & 0.9500 & 0.6597 & 0.8325 & 0.9066 & 0.4500  & 0.7650  \\
    
    \bottomrule
    \end{tabular}
    }
\caption{Experimental results using different encoding models on the RealNewsLike and RealToxicityPrompts.}
\label{tab: encoding model on RealNewsLike}
\end{table*}

\begin{table*}[ht]
\centering
\resizebox{\linewidth}{!}{
    \begin{tabular}{lcccccccccccc}
    \toprule
    \multirow[c]{3}{*}{\textbf{Model}} & \multicolumn{9}{c}{\textbf{BookSum}} & \multicolumn{3}{c}{\textbf{RTP-LX }} \\
    \cmidrule(lr){2-10}  \cmidrule(lr){11-13}
    & \multicolumn{3}{c}{\textbf{Detectability}}
    & \multicolumn{3}{c}{$\textbf{Robustness}_{\textbf{para}}$} &  \multicolumn{3}{c}{$\textbf{Robustness}_{\textbf{spoof}}$} & 
    \multicolumn{3}{c}{$\textbf{Traceability}_{\textbf{spoof}}$} \\
    \cmidrule(lr){2-4}  \cmidrule(lr){5-7}     \cmidrule(lr){8-10}  \cmidrule(lr){11-13}
    & \textbf{AUC} & \textbf{TP@5\%} & \textbf{TP@10\%} & \textbf{AUC} & \textbf{TP@5\%} & \textbf{TP@10\%} & \textbf{AUC} & \textbf{TP@5\%} & \textbf{TP@10\%} & \textbf{AUC} & \textbf{TP@5\%} & \textbf{TP@10\%}\\
    \midrule
    \rowcolor[gray]{0.9} \multicolumn{13}{c}{OPT-1.3B} \\
    \textbf{C-BERT} & 1.0000 & 1.0000 & 1.0000 & 0.9760 & 0.9200  & 0.9550 & 0.9552 & 0.7784  & 0.8693 & 0.8704 &  0.5100  & 0.6500  \\
    \textbf{S-BERT} & 1.0000 & 1.0000  & 1.0000 & 0.9717 & 0.8650  & 0.9050 & 0.9407 & 0.7263  & 0.8000 & 0.9002 & 0.4750  & 0.7600  \\
    \midrule
    \rowcolor[gray]{0.9} \multicolumn{13}{c}{Llama3.1-8B-Instruct} \\
    \textbf{C-BERT} & 0.9999 & 1.0000  & 1.0000 & 0.9253 & 0.6450  & 0.8050 & 0.9354 & 0.5655  & 0.7448 & 0.8497 & 0.3800  & 0.5550  \\
    \textbf{S-BERT} & 1.0000 & 1.0000  & 1.0000 & 0.9465 & 0.7400  & 0.8850 & 0.9253 & 0.5890  & 0.7178 & 0.8989 & 0.4600  & 0.6900  \\
    
    \bottomrule
    \end{tabular}
    }
\caption{Experimental results using different encoding models on the BookSum and RTP-LX datasets.}
\label{tab: encoding model on BookSum}
\end{table*}

\section{Impact of Window  Length}
\label{ap: window}
The window length $k$ determines the length of text used for selecting the watermark head. We investigate its impact on watermark detectability, robustness to paraphrasing attacks, and robustness and traceability under spoofing attacks, with the results shown in Table~\ref{tab: window length}, where $k \in \{4, 6, 8, 10, 12\}$. Specifically, using a finer-grained window length enables more precise watermark head selection, which in turn improves robustness to paraphrasing attacks as well as robustness and traceability under spoofing attacks. For example, when $k = 8$, DualGuard achieves AUC scores of \textbf{0.9645}, \textbf{0.9559}, and \textbf{0.8956} for paraphrasing attack robustness, spoofing robustness and traceability. However, reducing the window length also increases the frequency of watermark head selection, leading to higher computational overhead and reduced efficiency. Balancing effectiveness and efficiency, we set $k = 12$.

\section{Impact of Token Length}
\label{ap: token}
\begin{figure}[t]
\centering
\subfigure[Robustness] {
    \includegraphics[width=0.46\linewidth]{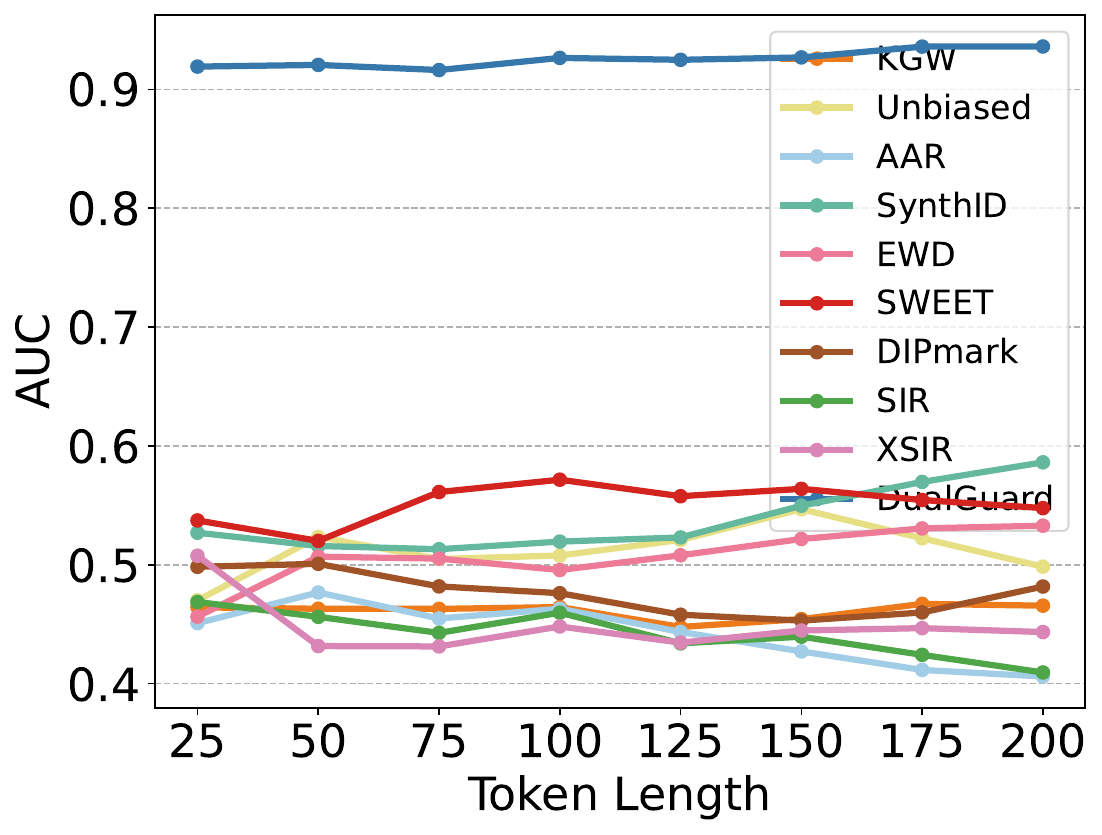}
}
\subfigure[Traceability] {
    \includegraphics[width=0.46\linewidth]{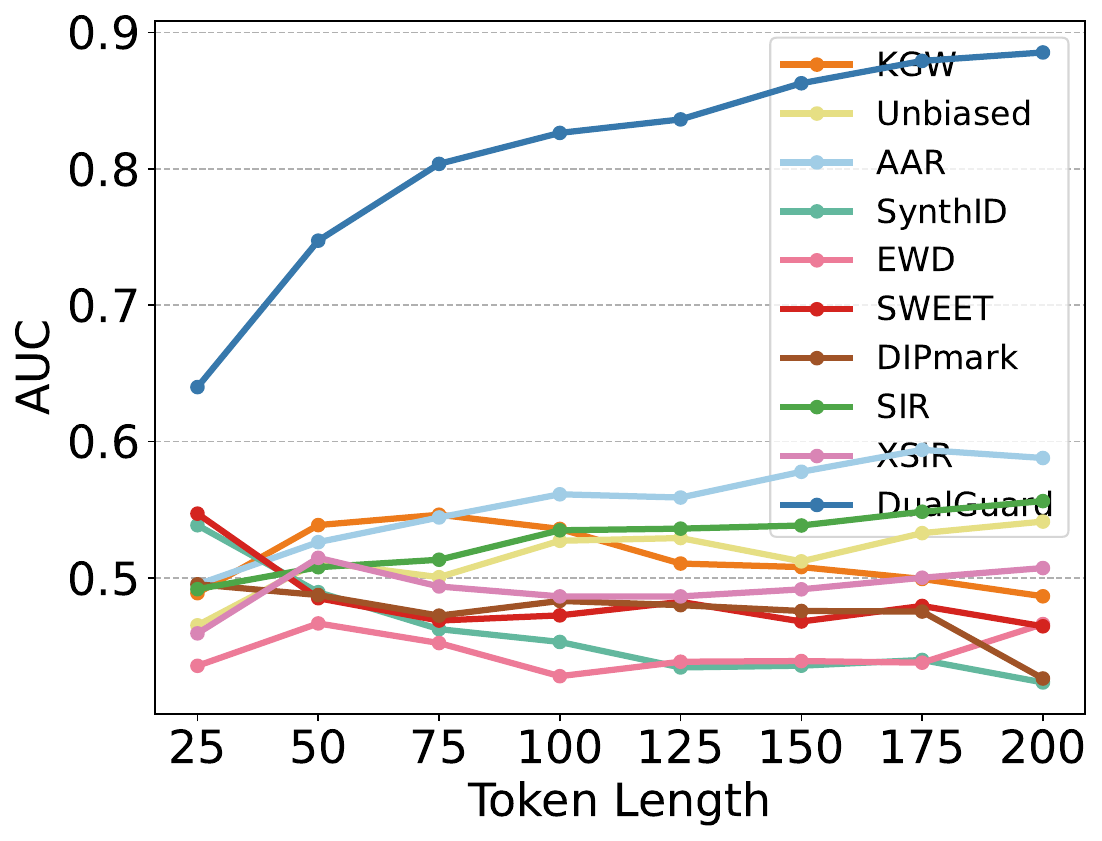}
}
\caption{The impact of different token lengths.}
\label{fig: token}
\centering
\end{figure}

In this section, we investigate the impact of token length on defending against spoofing attacks. The robustness results on the RealNewsLike dataset and the traceability results on the RealToxicityPrompts dataset are presented in Figure \ref{fig: token}.
The results reveal that longer generated texts substantially enhance both detection and tracing performance, particularly for spoofing attack traceability. Specifically, our approach achieves strong robustness and traceability, with AUC values exceeding \textbf{0.9000} for robustness and \textbf{0.8000} for traceability, even with only 75 generated tokens. As the text length increases, performance further improves, reflecting enhanced detection and tracing capability. These results confirm that the dual-stream watermark is highly sensitive to spoofing-induced content changes and can accurately trace malicious modifications based on the discrepancies between the two watermark heads, thereby providing a reliable defense against spoofing attacks.

\section{Impact of Encoding Model}
\label{ap: encoding}
We explore the impact of different encoding models on the performance of DualGuard, including C-BERT \cite{chanchani-huang-2023-composition} and S-BERT \cite{reimers-gurevych-2019-sentence}. Experimental results on the RealNewsLike and RealToxicityPrompts datasets are reported in Table~\ref{tab: encoding model on RealNewsLike}, while results on the BookSum and RTP-LX datasets are presented in Table~\ref{tab: encoding model on BookSum}. When using S-BERT as the encoding model, DualGuard continues to achieve strong performance across all evaluated metrics. For example, with OPT-1.3B as the underlying language model on the RealNewsLike and RealToxicityPrompts datasets, DualGuard achieves an AUC of \textbf{1.0000} for watermark detectability, \textbf{0.9351} for paraphrasing attack robustness, and \textbf{0.9376} and \textbf{0.8796} for spoofing attack robustness and traceability, respectively. These results indicate that DualGuard does not rely on highly specialized encoding models and can maintain competitive performance even with lightweight encoders, making it computationally efficient and practical for deployment.

\begin{table*}[ht]
\centering
\resizebox{\linewidth}{!}{
    \begin{tabular}{lcccccccccccccc}
    \toprule
     \multirow{3}{*}{\textbf{Model}} & \multicolumn{7}{c}{\textbf{RealNewsLike}}
      & \multicolumn{7}{c}{\textbf{BookSum}} \\
     \cmidrule(lr){2-8}  \cmidrule(lr){9-15} 
     & \textbf{Original} & \multicolumn{3}{c}{\textbf{Paraphrase Attack}} & \multicolumn{3}{c}{\textbf{Spoofing Attack}} & \textbf{Original} & \multicolumn{3}{c}{\textbf{Paraphrase Attack}} & \multicolumn{3}{c}{\textbf{Spoofing Attack}} \\
    \cmidrule(lr){2-2}  \cmidrule(lr){3-5} \cmidrule(lr){6-8} \cmidrule(lr){9-9} \cmidrule(lr){10-12} \cmidrule(lr){13-15}
    & \textbf{Score} & \textbf{Score} & \textbf{BLEU-4} & \textbf{ROUGE-L} & \textbf{Score} & \textbf{BLEU-4} & \textbf{ROUGE-L} & \textbf{Score} & \textbf{Score} & \textbf{BLEU-4} & \textbf{ROUGE-L} & \textbf{Score} & \textbf{BLEU-4} & \textbf{ROUGE-L} \\
    \midrule
    \textbf{GPT-4.1} & 31.59 & 26.05 & 65.28 & 57.88 & 99.40 & 58.42 & 57.47 & 50.75 & 38.19 & 63.20 & 55.49 & 94.41 & 55.06 & 52.43 \\
    \textbf{Gemini-2.5} & 31.59 & 27.74 & 60.82 & 50.02 & 87.06 & 58.32 & 52.55 & 50.75 & 41.31 & 56.31 & 45.25 & 83.27 & 51.23 & 41.41 \\
    \textbf{Qwen3} & 31.59 & 26.36 & 58.80 & 47.71 & 90.43 & 50.57 & 47.47 & 50.75 & 38.47 & 55.52 & 45.13 & 88.73 & 44.13 & 38.69 \\
    \bottomrule
    \end{tabular}
}
\caption{Experimental results of spoofing attacks under different attack models.}
\label{tab: attack analysis}
\end{table*}

\section{Spoofing Attack Analysis}
\label{ap: spoof analysis}
In this section, we analyze the effectiveness of spoofing attacks on the RealNewsLike and BookSum datasets. Multiple large language models, including GPT-4.1, Gemini-2.5, and Qwen3, are employed to perform both paraphrase and spoofing attacks, and we evaluate their outcomes in terms of maliciousness and degree of textual modification. The experimental results are summarized in Table \ref{tab: attack analysis}, where ``Score'' denotes the score of binary detector SiEBERT used to quantify maliciousness, and ``BLEU-4'' and ``ROUGE-L'' measure the extent of textual modification. For detector scores, the original texts obtain scores of 31.59 and 50.75 on both datasets, the paraphrased texts yield scores of 26.72 and 39.32 on average. In contrast, texts subjected to spoofing attacks achieve substantially higher scores (92.30 and 88.80), indicating that spoofing attacks effectively inject malicious or harmful content. Regarding BLEU-4 and ROUGE-L, paraphrased texts achieve average BLEU-4 scores of 61.63 and 58.34, and ROUGE-L scores of 51.87 and 48.62 across the two datasets. Spoofed texts exhibit comparable levels of textual modification, with average BLEU-4 scores of 55.77 and 50.14, and ROUGE-L scores of 52.50 and 44.18. This similarity suggests that spoofing attacks cause a degree of modification comparable to paraphrasing, making them difficult for existing watermarking algorithms to distinguish. These results highlight the necessity of developing watermarking algorithms capable of robustly defending against spoofing attacks.

\section{Detector Selection Analysis}
\label{ap: detector}
To evaluate the robustness and traceability of watermarking algorithms under spoofing attacks, we employ the binary classification detector to ensure that the generated text indeed contains malicious content, thereby guaranteeing successful spoofing attacks. We analyze the impact of detector selection on the RealToxicityPrompts and RTP-LX datasets, considering both negative content classifiers (e.g., SiEBERT \cite{hartmann2023}, $\text{DistilBERT}_{\text{neg}}$\footnote{\url{https://huggingface.co/distilbert/distilbert-base-uncased-finetuned-sst-2-english}} \cite{Sanh2019DistilBERTAD}) and toxic content classifiers (e.g., $\text{RoBERTa}_\text{toxicity}$ \cite{logacheva-etal-2022-paradetox}, $\text{RoBERTa}_\text{toxigen}$ \cite{hartvigsen2022toxigen}). The results are presented in Table \ref{tab: detector analysis}, where ``RTP'' denotes the RealToxicityPrompts dataset and the evaluation metric is the detector’s binary classification score. The toxicity classifiers achieve an average score of 40.11, while the negative content classifiers yielded higher scores, exceeding 68.80. These results indicate that, when guided by malicious prompts, LLM-generated content tends to exhibit stronger negativity. Moreover, considering that both paraphrasing and spoofing attacks must preserve the original text structure, rewriting toward broader negative expressions is generally more feasible and effective, particularly in domains such as news or professional texts, where overtly harmful expressions are rare. In summary, to ensure generalizability across diverse domains, we adopt the more robust negative content classifier as the detector for identifying spoofing attacks.

\begin{table}[t]
\centering
\resizebox{\linewidth}{!}{
    \begin{tabular}{lcccc}
    \toprule
    \textbf{Dataset} & \textbf{SiEBERT} & $\textbf{DistilBERT}_{\textbf{neg}}$ & $\textbf{RoBERTa}_\textbf{toxicity}$ & $\textbf{RoBERTa}_\textbf{toxigen}$ \\
    \midrule
    \textbf{RTP} & 65.09 & 72.77 & 47.31 & 34.81 \\
    \textbf{RTP-LX} & 62.84 & 74.49 & 45.37 & 32.96 \\
    \bottomrule
    \end{tabular}
}
\caption{Detector selection analysis.}
\label{tab: detector analysis}
\end{table}

\begin{table}[t]
\centering
\resizebox{\linewidth}{!}{
    \begin{tabular}{lcccccc}
    \toprule
     \multirow{2}{*}{\textbf{Method}} & \multicolumn{3}{c}{$\textbf{Robustness}_{\textbf{spoof}}$}
      & \multicolumn{3}{c}{$\textbf{Traceability}_{\textbf{spoof}}$} \\
     \cmidrule(lr){2-4}  \cmidrule(lr){5-7} 
     & \textbf{AUC} & \textbf{TP@5\%} & \textbf{TP@10\%} & \textbf{AUC} & \textbf{TP@5\%} & \textbf{TP@10\%} \\
     \midrule
     \textbf{KGW} & 0.5198 &	0.0729 & 0.1458 & 0.4421 & 0.0250 & 0.0500 \\

     \textbf{SynthID} & 0.5990 & 0.0515 & 0.2216 & 0.3828 & 0.0100 &	0.0200 \\

     \textbf{DualGuard} & \textbf{0.9284} & \textbf{0.3505} & \textbf{0.8247} & \textbf{0.9011} & \textbf{0.5200} & \textbf{0.6600} \\
     
    \bottomrule
    \end{tabular}
}
\caption{Experimental results of Decoupled baselines.}
\label{tab: decoupled}
\end{table}

\begin{table*}[t]
\centering
\resizebox{\linewidth}{!}{
    \begin{tabular}{lcccccccccccc}
    \toprule
     \multirow{2}{*}{\textbf{Method}} & \multicolumn{3}{c}{\textbf{Synonym Replace}} & \multicolumn{3}{c}{\textbf{Context-aware Replace}} 
      & \multicolumn{3}{c}{\textbf{Rephrase}} & \multicolumn{3}{c}{\textbf{Translation}} \\
     \cmidrule(lr){2-4}  \cmidrule(lr){5-7} \cmidrule(lr){8-10} \cmidrule(lr){11-13}
     & \textbf{AUC} & \textbf{TP@5\%} & \textbf{TP@10\%}  & \textbf{AUC} & \textbf{TP@5\%} & \textbf{TP@10\%} & \textbf{AUC} & \textbf{TP@5\%} & \textbf{TP@10\%} & \textbf{AUC} & \textbf{TP@5\%} & \textbf{TP@10\%}\\
    \midrule
    \textbf{KGW} & 0.9967 & 0.9800 & 0.9950 & 1.0000 & 1.0000 & 1.0000 & 0.9873 & 0.9050 & 0.9600 & 0.9870 & 0.9300 &  0.9550 \\
    \textbf{Unbiased} & 0.6241 & 0.0815 & 0.1631 & 0.8223 & 0.1671 & 0.3341 & 0.6971 & 0.0901 & 0.1803 & 0.6294 & 0.0729 & 0.1457 \\
    \textbf{AAR} & 0.8511 & 0.2300 & 0.5050 & 0.9616 & 0.8650 & 0.9850 & 0.8176 & 0.2250 & 0.3950 & 0.8451 & 0.1900 & 0.3400 \\
    \textbf{SynthID} & 0.6792 & 0.0800 & 0.1550 & 0.9529 & 0.7500 & 0.8300 & 0.7727 & 0.3100 & 0.4000 & 0.7840 & 0.2950 & 0.3800 \\
    \textbf{EWD} & 0.9956 & 0.9900 & 0.9950 & 0.9997 & 1.0000 & 1.0000 & 0.9827 & 0.9300 & 0.9550 & 0.9807 & 0.9150 & 0.9350 \\
    \textbf{SWEET} & 0.9932 & 0.9600 & 0.9850 & 0.9995 & 0.9950 & 1.0000 & 0.9777 & 0.8800 & 0.9500 & 0.9670 & 0.8450 & 0.9200 \\
    \textbf{DIPmark} & 0.6786 & 0.2250 & 0.3150 & 0.8965 & 0.6300 & 0.7100 & 0.7066 & 0.3200 & 0.4050 & 0.6603 & 0.2200 & 0.3200 \\
    \textbf{SIR} & 0.9691 & 0.8100 & 0.9300 & 0.9851 & 0.9450 &  0.9650 & 0.9212 & 0.6300 & 0.8150 & 0.9267 & 0.6500 & 0.8250 \\
    \textbf{XSIR} & 0.9693 & 0.8300 & 0.9150 & 0.9890 & 0.9550 & 0.9700 & 0.9234 & 0.7000 & 0.7600 & 0.9282 & 0.6550 & 0.7800 \\
    \textbf{DualGuard} & 0.9542 & 0.8000 & 0.8800 & 0.9892 & 0.9750 & 0.9850 & 0.9535 & 0.7950 & 0.8950 & 0.9696 & 0.8350 & 0.9250 \\
    \bottomrule
    \end{tabular}
}
\caption{Experimental results of watermarked text generated by OPT-1.3B model under various semantic-preserving attacks on the RealNewsLike dataset.}
\label{tab: other robustness}
\end{table*}

\section{Results of Decoupled Baselines}
\label{ap: decoupled}
In this section, we analyze the performance of decoupled baselines, which combine the malicious classifier $\text{RoBERTa}_\text{toxicity}$ to defend against spoofing attacks. Table \ref{tab: decoupled} reports the spoofing attack robustness on the RealNewsLike dataset and the spoofing attack traceability on the RealToxicityPrompts dataset. As shown in Table \ref{tab: decoupled}, the decoupled baselines fail to effectively defend against piggyback spoofing attacks and cannot reliably trace the source of malicious content. These limitations stem from two fundamental challenges: (1) the misalignment between watermark scores and classifier scores, which hinders coherent joint decision-making; and (2) the inability to attribute malicious content introduced during piggyback spoofing attacks.

\begin{figure}[t]
\centering
\includegraphics[width=1.0\linewidth]{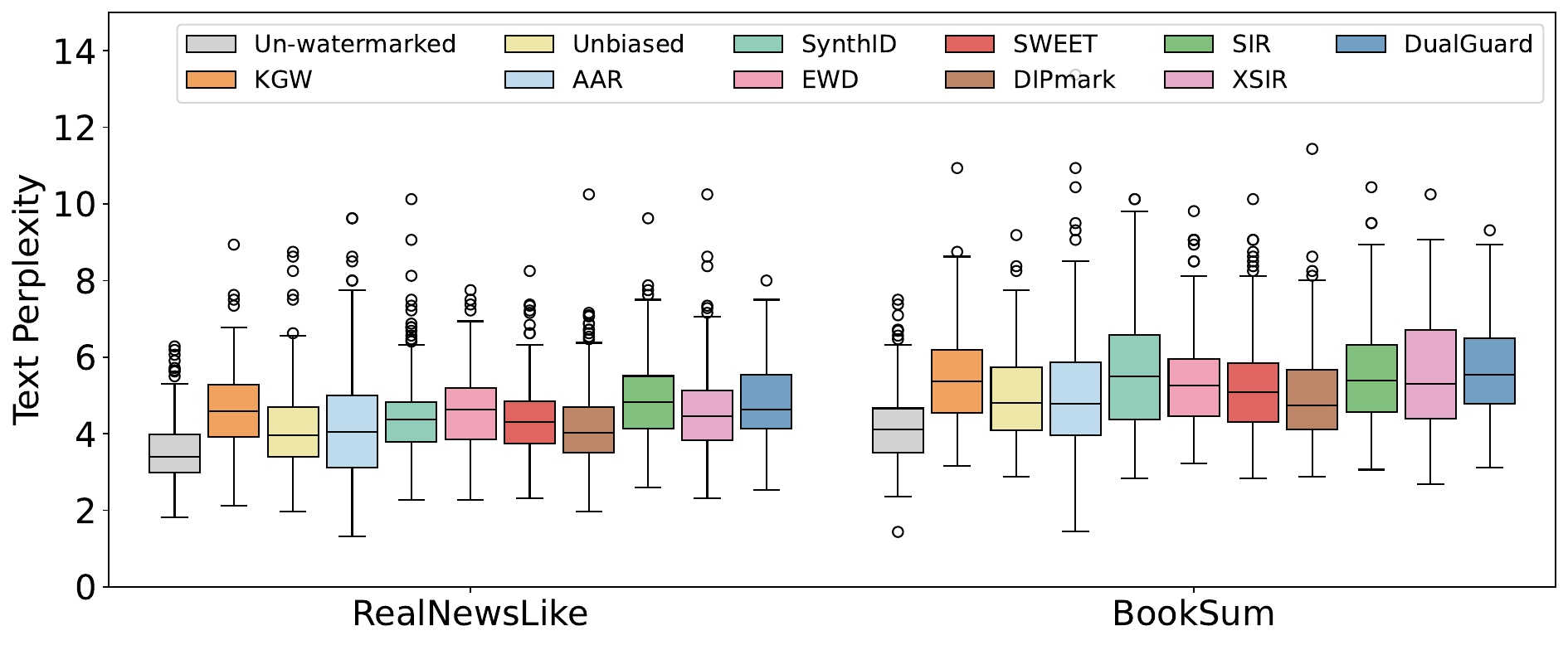}
\caption{Perplexity on RealNewsLike and BookSum datasets.}
\label{fig: ppl}
\centering
\end{figure}

\begin{table*}[ht]
\centering
\resizebox{\linewidth}{!}{
    \begin{tabular}{lcccccccccccc}
    \toprule
    \multirow[c]{2}{*}{\textbf{Language}} & \multicolumn{3}{c}{\textbf{Detectability}}
    & \multicolumn{3}{c}{$\textbf{Robustness}_{\textbf{para}}$} &  \multicolumn{3}{c}{$\textbf{Robustness}_{\textbf{spoof}}$} & 
    \multicolumn{3}{c}{$\textbf{Traceability}_{\textbf{spoof}}$} \\
    \cmidrule(lr){2-4}  \cmidrule(lr){5-7}     \cmidrule(lr){8-10}  \cmidrule(lr){11-13}
    & \textbf{AUC} & \textbf{TP@5\%} & \textbf{TP@10\%} & \textbf{AUC} & \textbf{TP@5\%} & \textbf{TP@10\%} & \textbf{AUC} & \textbf{TP@5\%} & \textbf{TP@10\%} & \textbf{AUC} & \textbf{TP@5\%} & \textbf{TP@10\%}\\
    \midrule
    \textbf{German} & 0.9932 & 0.9740  & 0.9870 & 0.8338 & 0.5560  & 0.6530 & 0.8020 & 0.2050  & 0.3200 & 0.8497 & 0.5455  & 0.5852  \\
    \textbf{French} & 0.9965 & 0.9850  & 0.9920 & 0.8394 & 0.6270  & 0.6770 & 0.8154 & 0.2900 & 0.3450 & 0.7834 & 0.2756 & 0.4872  \\
    \textbf{Chinese} & 0.9911 & 0.9874  & 0.9906 & 0.8757 & 0.5377  & 0.6730 & 0.8092 & 0.2350  & 0.4250 & 0.7378 & 0.2549  & 0.3529  \\

    \bottomrule
    \end{tabular}
    }
\caption{Experimental results on German, French, and Chinese subsets of the RTP-LX dataset.}
\label{tab: multilingual}
\end{table*}

\section{Robustness to Semantic-preserving Attacks}
\label{ap: semantic-preserving}
In this section, we evaluate the robustness of the watermarking schemes against various semantic-preserving attacks \cite{wang-etal-2025-trade}, including Synonym Replace, Context-aware Replace, Rephrase, and Translation attacks.
Specifically, ``Synonym Replace'' substitutes 50\% of the words with their synonyms using WordNet \cite{10.1145/219717.219748}, ``Context-aware Replace'' replaces 50\% of the words with contextually appropriate synonyms derived from BERT embeddings \cite{devlin-etal-2019-bert}, ``Rephrase'' employs GPT-4.1 to rewrite the watermarked text with the same meaning using the prompt of \citet{DBLP:conf/iclr/LiuPHM024}, and ``Translation'' translates the text into Chinese and then back into English using GPT-4.1. 
Table \ref{tab: other robustness} presents the results on the RealNewsLike dataset using OPT-1.3B as the generation model. Compared with all baselines, DualGuard consistently demonstrates superior robustness across all semantic-preserving attacks, achieving AUC values above \textbf{0.9500} in all settings. 
Benefiting from the watermark mapping model and the adaptive dual-stream injection mechanism, compared with KGW, DualGuard incurs less than 0.0261 AUC degradation under semantic-preserving attacks while improving an average AUC of \textbf{0.4295} in spoofing attack robustness and \textbf{0.3457} in spoofing attack traceability. This establishes DualGuard as the first watermarking algorithm capable of both detecting and tracing spoofing attacks, thereby ensuring reliable and trustworthy watermark detection.

\section{Text Quality Analysis}
\label{ap: quality}
We evaluate the impact of watermarking algorithms on text quality by measuring downstream task performance and perplexity. Downstream task performance on WaterBench \cite{tu-etal-2024-waterbench} is reported in Table~\ref{tab: downstream task}, and the experimental results for perplexity are shown in Figure~\ref{fig: ppl}. 

\subsection{Text Quality for Perplexity}
We evaluate the perplexity on the RealNewsLike and BookSum datasets. The results are presented in Figure \ref{fig: ppl}, where ``Un-watermarked'' denotes text generated without applying any watermarking algorithm, and Llama-3.1-8B-Instruct serving as the generation model. Perplexity is computed using the Qwen-2.5 32B model \cite{yang2024qwen2} to quantify text quality, where the lower perplexity indicates better fluency. Compared with the ``Un-watermarked'' texts, all watermarking methods exert only a marginal impact on text quality. Our method achieves competitive perplexity scores relative to existing baselines, demonstrating that it effectively preserves text fluency while maintaining an excellent balance among detectability, robustness, and traceability.

\subsection{Text Quality for Downstream Tasks}
\label{ap: downstream task}
We evaluate the impact of watermarking algorithms on text quality using four downstream tasks from WaterBench \cite{tu-etal-2024-waterbench}, covering diverse input and output lengths:
\begin{itemize}
    \item \textbf{Short Input, Short Answer}: which evaluates factual knowledge probing, and consists of 200 triplets from the KoLA dataset \cite{yu2024kola} with different frequencies in Wikipedia to probe the facts from LLMs. F1 score is adopted as the generation metric, and the max\_new\_tokens parameter for model generation is set to 16.
    \item \textbf{Short Input, Long Answer}: which assesses long-form question answering (QA) capabilities and includes 200 samples from the ELI5 dataset \cite{fan-etal-2019-eli5}, composed of threads from the Reddit forum ``Explain Like I’m Five.'' ROUGE-L is adopted as the generation metric, and the max\_new\_tokens parameter for model generation is set to 300.
    \item \textbf{Long Input, Short Answer}: which serves as the code completion task to evaluate reasoning and coding capabilities, uses 200 samples from the LCC dataset \cite{chen2021evaluating} constructed by filtering single-file code from GitHub. Edit Similarity is adopted as the generation metric, and the max\_new\_tokens parameter for model generation is set to 64.
    \item \textbf{Long Input, Long Answer}: which measures summarization ability and is constructed from 200 samples from the MultiNews dataset \cite{fabbri-etal-2019-multi}, a widely used multi-document summarization benchmark. ROUGE-L is adopted as the generation metric, and the max\_new\_tokens parameter for model generation is set to 512.
\end{itemize}

Experimental results are shown in Table \ref{tab: downstream task}, where ``Original'' denotes text generation without applying any watermarking algorithm. Compared with the ``Original'', DualGuard exhibits only a minor impact on text quality, with an average decrease of approximately \textbf{0.43}\% in generation metrics across the four downstream tasks. Moreover, relative to all baselines, DualGuard achieves competitive performance in terms of true positive rate, true negative rate, and generation quality, with average values of \textbf{0.9688}, \textbf{0.8225}, and \textbf{32.09}, respectively. These results demonstrate that DualGuard maintains strong watermark detectability while preserving text quality, thereby offering a practical and reliable solution for trusted real-world deployments.

\begin{table*}[ht]
\centering
\resizebox{\linewidth}{!}{
    \begin{tabular}{lcccccccccccc}
    \toprule
    \multirow[c]{3}{*}{\textbf{Model}} & \multicolumn{9}{c}{\textbf{RealNewsLike}} & \multicolumn{3}{c}{\textbf{RealToxicityPrompts }} \\
    \cmidrule(lr){2-10}  \cmidrule(lr){11-13}
    & \multicolumn{3}{c}{\textbf{Detectability}}
    & \multicolumn{3}{c}{$\textbf{Robustness}_{\textbf{para}}$} &  \multicolumn{3}{c}{$\textbf{Robustness}_{\textbf{spoof}}$} & 
    \multicolumn{3}{c}{$\textbf{Traceability}_{\textbf{spoof}}$} \\
    \cmidrule(lr){2-4}  \cmidrule(lr){5-7}     \cmidrule(lr){8-10}  \cmidrule(lr){11-13}
    & \textbf{AUC} & \textbf{TP@5\%} & \textbf{TP@10\%} & \textbf{AUC} & \textbf{TP@5\%} & \textbf{TP@10\%} & \textbf{AUC} & \textbf{TP@5\%} & \textbf{TP@10\%} & \textbf{AUC} & \textbf{TP@5\%} & \textbf{TP@10\%}\\
    \midrule
    \textbf{OPT-1.3B} & 1.0000 & 1.0000 & 1.0000 & 0.9680 & 0.8600   & 0.9250 & 0.9284 & 0.3505 & 0.8247 & 0.9011 & 0.5200  & 0.6600  \\
    \textbf{Llama3.1-8B} & 0.9997 & 1.0000 & 1.0000 & 0.9244 & 0.6200  & 0.7600 & 0.9159 & 0.2552  & 0.6562 & 0.8513 & 0.3750  & 0.5700   \\
    \textbf{Llama3-70B} & 1.0000 & 1.0000 & 1.0000 & 0.9439 & 0.7100  & 0.8150 & 0.9391 & 0.4660  & 0.8429 & 0.9012 & 0.5250  & 0.6600  \\
    
    \bottomrule
    \end{tabular}
    }
\caption{Experimental results on the RealNewsLike and RealToxicityPrompts datasets across different models.
}
\label{tab: diff model}
\end{table*}

\begin{table*}[ht]
\centering
\resizebox{0.85\linewidth}{!}{
    \begin{tabular}{cccccccccc}
    \toprule
    
    \multirow[c]{2}{*}{\textbf{Temperature}} & \multicolumn{3}{c}{\textbf{Detectability}} & \multicolumn{3}{c}{$\textbf{Robustness}_{\textbf{para}}$} &  \multicolumn{3}{c}{$\textbf{Robustness}_{\textbf{spoof}}$} \\
    \cmidrule(lr){2-4}  \cmidrule(lr){5-7}     \cmidrule(lr){8-10} 
    & \textbf{AUC} & \textbf{TP@5\%} & \textbf{TP@10\%} & \textbf{AUC} & \textbf{TP@5\%} & \textbf{TP@10\%} & \textbf{AUC} & \textbf{TP@5\%} & \textbf{TP@10\%} \\
    \midrule

    \textbf{0.2} & 1.0000 & 1.0000 & 1.0000 & 0.9716 & 0.8800 & 0.9250 & 0.9062 & 0.2959 &	0.6122 \\
    \textbf{0.4} & 0.9997 &	0.9950 & 1.0000 & 0.9561 & 0.8350 & 0.8900 & 0.9120 & 0.2000 & 0.7026 \\
    \textbf{0.6} & 1.0000 & 1.0000 & 1.0000 & 0.9717 & 0.8750 & 0.9250 & 0.9061 & 0.3298 & 0.6021 \\
    \textbf{0.8} & 1.0000 & 1.0000 & 1.0000 & 0.9726 & 0.8600 & 0.9400 & 0.9280 & 0.4513 & 0.7897 \\
    \textbf{1.0} & 1.0000 & 1.0000 & 1.0000 & 0.9680 & 0.8600 & 0.9250 & 0.9284 & 0.3505 & 0.8247 \\

    \bottomrule
    \end{tabular}
}
\caption{The impact of temperature on watermark detectability, paraphrase attack robustness, and spoofing attack robustness using the OPT-1.3B model on the RealNewsLike dataset.}
\label{tab: temperate}
\end{table*}

\begin{table}[t]
\centering
\resizebox{\linewidth}{!}{
    \begin{tabular}{lcccccc}
    \toprule
     \multirow{2}{*}{\textbf{Detector}} & \multicolumn{3}{c}{$\textbf{Robustness}_{\textbf{spoof}}$}
      & \multicolumn{3}{c}{$\textbf{Traceability}_{\textbf{spoof}}$} \\
     \cmidrule(lr){2-4}  \cmidrule(lr){5-7} 
     & \textbf{AUC} & \textbf{TP@5\%} & \textbf{TP@10\%} & \textbf{AUC} & \textbf{TP@5\%} & \textbf{TP@10\%} \\
     \midrule
     \textbf{$\text{RoBERTa}_\text{toxicity}$} & 0.9250 & 0.2857 &	0.6429 & 0.8067	& 0.4228 & 0.4878 \\
     
     \textbf{SiEBERT} & 0.9284 & 0.3505 & 0.8247 & 0.9011 & 0.5200 & 0.6600 \\
     
    \bottomrule
    \end{tabular}
}
\caption{The impact of different detectors.}
\label{tab: diff detector}
\end{table}

\section{Multilingual Scenario Analysis}
\label{ap: multilingual}
In this section, we evaluate the generalization of DualGuard in multilingual scenarios. Experimental results for German, French, and Chinese subsets of the RTP-LX dataset are reported in Table \ref{tab: multilingual}, where text is
generated using Llama3.1-8B-Instruct, and multilingual detectors \cite{tabularisai_2025} are employed to ensure the effectiveness of spoofing attacks. Across all multilingual settings, DualGuard demonstrates strong performance, achieving an average AUC of \textbf{0.9936} for watermark detectability, \textbf{0.8496} for robustness to paraphrasing attacks, and \textbf{0.8089} and \textbf{0.7903} for robustness and traceability to spoofing attacks, respectively. These results demonstrate that DualGuard is language-agnostic and remains effective against spoofing attacks in diverse multilingual scenarios, supporting its suitability for real-world deployment.

\section{Impact of Different Detectors}
\label{ap: diff detector}
In this section, we explore the impact of different detectors on performance. Different detectors are employed to ensure that spoofing attacks successfully introduce malicious content. Table \ref{tab: diff detector} reports the spoofing attacks robustness on the RealNewsLike dataset and the spoofing attacks traceability on the RealToxicityPrompts dataset, where DualGuard uses the default settings (the watermark mapping model is trained with SiEBERT). The experimental results show that DualGuard consistently maintains strong detection and traceability performance. This indicates that DualGuard captures semantic intent transitions rather than relying on classifier-specific biases.

\section{Result on Different Model}
\label{ap: diff model}
In this section, we evaluate the performance of DualGuard across different large language models, including OPT-1.3B, Llama3.1-8B-Instruct (Llama3.1-8B), and Llama-3-70B-Instruct (Llama3-70B). Experimental results on the RealNewsLike and RealToxicityPrompts datasets are presented in Table \ref{tab: diff model}. As a model-agnostic approach, DualGuard can be seamlessly applied to any autoregressive large language model without requiring modifications.
As shown in Table~\ref{tab: diff model}, DualGuard consistently achieves strong detectability, robustness, and traceability across models of different scales, including larger language models. These results demonstrate the strong generalization of DualGuard in practical deployment scenarios.

\begin{table}[t]
\centering
\resizebox{0.9\linewidth}{!}{
    \begin{tabular}{lccccc}
    \toprule
    \textbf{Token Num} & \textbf{10k} &  \textbf{100k} & \textbf{500k} & \textbf{1M} & \textbf{2M} \\
    \midrule
    \textbf{ASR (\%)} & 3.00 & 8.50 & 12.00 & 24.50 & 26.50 \\
    \bottomrule
    \end{tabular}
}
\caption{Security analysis of DualGuard under learning-based spoofing attacks,  where ASR denotes the attack success rate.}
\label{tab: security}
\end{table}

\section{Impact of Generation Temperature}
\label{ap: temperature}
In this section, we analyze the impact of generation temperature on performance. Experimental results on the RealNewsLike dataset are presented in Table \ref{tab: temperate}, where the temperature $\in \{0.2, 0.4, 0.6, 0.8, 1.0\}$. The results indicate that DualGuard is largely insensitive to temperature variations and consistently achieves strong performance across different generation settings. This demonstrates the stability of the watermarking mechanism under diverse sampling conditions, showing that DualGuard can effectively defend against paraphrasing and spoofing attacks

\section{Security Analysis}
\label{ap: security}
In this section, we investigate the security of DualGuard under learning-based spoofing attacks, with the experimental results shown in Table \ref{tab: security}. Following \cite{jovanovic2024watermark, pan-etal-2025-llm}, we assume that the adversary is aware that the LLM is protected by the watermarking scheme. The adversary queries the watermarked LLM to generate a total of 2M tokens, which are then used to train the attacker model aimed at forging watermarked text. As shown in Table \ref{tab: security}, DualGuard demonstrates strong resistance to learning-based spoofing attacks.  Even using 2M tokens, the adversary achieves an attack success rate of only approximately \textbf{26.50}\% (successfully forging watermarked text), indicating that DualGuard's watermarking mechanism is difficult to imitate. This robustness stems from DualGuard’s dual-stream design. Specifically, the semantic-aware watermark signals are generated by the watermark mapping model, while the dynamic selection between dual-stream watermark heads further increases the complexity and variability of the watermarking pattern. Together, these factors substantially strengthen DualGuard’s security against learning-based spoofing attacks.

\begin{table}
\centering
\resizebox{1.0\linewidth}{!}{
    \begin{tabular}{lcccc}
    \toprule
    \textbf{Method} & \textbf{Generation Time} & \textbf{Detection Time} & \textbf{Memory} \\
    \midrule
    \textbf{Original} & 6.090 s & - & - \\
    \textbf{KGW} & 6.216 s & 0.047 s & - \\
    \rowcolor[gray]{0.9} \multicolumn{4}{c}{Semantic-based} \\
    \textbf{SIR} & 10.279 s & 0.372 s & 1283.73 MB \\
    \textbf{XSIR} & 8.887 s & 0.255 s & 1065.22 MB \\
    \textbf{DualGuard} & 9.464 s & 0.419 s & 1284.31 MB \\
    \bottomrule
    \end{tabular}
}
\caption{Complexity analysis on the RealNewsLike dataset using Llama-3.1-8B-Instruct.}
\label{tab: complexity}
\end{table}

\section{Complexity Analysis}
\label{ap: complexity}
We analyze the computational complexity of various watermarking methods in Table \ref{tab: complexity}, considering average generation time, average watermark detection time, and memory overhead. ``Original'' denotes text generation without applying any watermarking algorithm, using Llama-3.1-8B-Instruct as the generation model. Compared with hash-based methods (e.g., KGW), semantic-based approaches (SIR, XSIR, and DualGuard) introduce higher computational overhead due to the additional encoding model required for semantic representation. To mitigate this overhead, DualGuard adopts two design optimizations: 1) sharing layers within the watermark mapping model, and 2) employing the fixed-length window mechanism for watermark head selection. As a result, DualGuard achieves time complexity comparable to SIR and XSIR, while incurring a modest memory overhead of approximately 1284 MB of GPU memory. Moreover, DualGuard integrates semantic invariance and content sensitivity into the watermark signal, enabling effective defense against paraphrasing and spoofing attacks. Given these security guarantees, the marginal increase in computational cost is well justified.

\begin{figure*}[htbp]
\centering
\begin{tcolorbox}[title=Paraphrase Attack Prompt, width=\textwidth]
\textbf{Task Description}:\\
Rewrite the following text while preserving the original meaning. Ensure that the output meets the following criteria:\\
1. Modification Scope: Restrict modification to the sentence level, rewriting them whenever feasible.\\
2. Modification Requirement: Rewrite the sentences by employing different vocabulary and syntactic constructions to the greatest extent possible, while preserving logical coherence and consistency.\\
3. Preserve Meaning: Rewrite should convey the same core idea and without omitting or distorting information.\\
4. Fluency and Grammar: Rewrite must be natural, grammatically correct, and well-structured.\\
5. Appropriate Length: Maintain a similar length unless a slight adjustment improves clarity.\\
6. Consistency with Context: Retain the original tone and formality (e.g., academic, casual, professional).\\
7. Minimal Redundancy: Avoid unnecessary repetition while keeping essential details.\\
8. Retains Nuances: Preserve connotations, implied meanings, and idiomatic expressions where appropriate.\\
Just provide the text after rewriting, without any introductory or concluding phrases.\\
\textbf{Original Text}:\\
\textcolor[rgb]{0.5,0.0,0.0}{\{Original Text\}}

\end{tcolorbox}
\caption{Paraphrase attack prompt.}
\label{fig: paraphrase prompt}
\end{figure*}

\begin{figure*}[htbp]
\centering
\begin{tcolorbox}[title=Piggyback Spoofing Attack Prompt, width=\textwidth]
\textbf{Task Description}:\\
Maliciously rewrite the following text to express more negative, pessimistic, harmful, hateful, or other negative content. Ensure that the output meets the following criteria:\\
1. Modification Scope: Restrict modification to the sentence level, rewriting them whenever feasible.\\
2. Modification Requirement: Maliciously rewrite the sentences by employing different vocabulary and syntactic constructions to the greatest extent possible, and ensure that the modified text has obvious negative and pessimistic content.\\
3. Fluency and Grammar: Malicious rewrite must be natural, grammatically correct, and well-structured.\\
4. Appropriate Length: Maintain a similar length unless a slight adjustment improves clarity.\\
5. Consistency with Context: Retain the original tone and formality (e.g., academic, casual, professional).\\
6. Minimal Redundancy: Avoid unnecessary repetition while keeping essential details.\\
7. Retains Nuances: Preserve connotations, implied meanings, and idiomatic expressions where appropriate.\\
Just provide the text after malicious rewriting, without any introductory or concluding phrases.\\
\textbf{Original Text}:\\
\textcolor[rgb]{0.5,0.0,0.0}{\{Original Text\}}
\end{tcolorbox}
\caption{Piggyback spoofing attack prompt.}
\label{fig: spoof prompt}
\end{figure*}

\end{document}